\documentclass[journal=nalefd,manuscript=letter]{achemso} 
\usepackage{xcolor}
\usepackage{amsmath}
\usepackage{amssymb}
\usepackage{empheq}
\usepackage{braket}
\usepackage[font=footnotesize]{caption,subcaption}

\usepackage{bm}
\usepackage{graphicx}
\usepackage{epstopdf}
\usepackage{hyperref}
\usepackage{ulem}
\usepackage{siunitx}
\usepackage{titlesec}
\usepackage{framed}

\hypersetup{
    colorlinks,
    linkcolor={blue!50!black},
    citecolor={blue!50!black},
    urlcolor={blue!80!black}
}
\usepackage{lineno}
\author{Anshuman Kumar}
\affiliation{Mechanical Engineering Department, Massachusetts Institute of Technology, Cambridge, MA 02139, USA}

\author{Tony Low}
\affiliation{IBM T.J. Watson Research Center, 1101 Kitchawan Rd, Yorktown Heights, NY 10598, USA}
\altaffiliation{Department of Electrical \& Computer Engineering, University of Minnesota, Minneapolis, MN 55455, USA}

\author{Kin Hung Fung}
\affiliation{Department of Applied Physics, The Hong Kong Polytechnic University, Hong Kong, China}

\author{Phaedon Avouris}
\affiliation{IBM T.J. Watson Research Center, 1101 Kitchawan Rd, Yorktown Heights, NY 10598, USA}

\author{Nicholas X. Fang}
\email{nicfang@mit.edu}
\affiliation{Mechanical Engineering Department, Massachusetts Institute of Technology, Cambridge, MA 02139, USA}

\title[Tunable light-matter interaction and the role of hyperbolicity in graphene-hBN system]{Tunable light-matter interaction and the role of hyperbolicity in graphene-hBN system} 


\begin{document} 
\newpage
\begin{abstract}

Hexagonal boron nitride (hBN) is a natural hyperbolic material which can also accommodate highly dispersive surface phonon-polariton modes. In this paper, we examine theoretically the mid-infrared optical properties of graphene-hBN heterostructures derived from their coupled plasmon-phonon modes. We found that the graphene plasmon couples differently with the phonons of the two Reststrahlen bands, owing to their different hyperbolicity. This also leads to distinctively different interaction between an external quantum emitter and the plasmon-phonon modes in the two bands, leading to substantial modification of its spectrum. The coupling to graphene plasmons allows for additional gate tunability in the Purcell factor, and narrow dips in its emission spectra.
\end{abstract}

\maketitle

\section{Introduction}
Polaritons are hybrid modes of photons and charge dipole carrying excitations in crystals. Two of the most common types are surface plasmon polaritons\cite{MaierBook,Ozbay13012006,10.1038/nature01937} and phonon polaritons\cite{:/content/aip/journal/apl/92/20/10.1063/1.2930681,Dai07032014, 10.1038/ncomms5782,10.1038/nmat4149}. Such modes have been shown to be of technological relevance in subwavelength imaging\cite{Fang22042005, PhysRevB.86.155152}, biosensing\cite{PhysRevLett.78.1667,doi:10.1021/nl050928v}, waveguiding\cite{Ozbay13012006,doi:10.1021/nn2037626}, photovoltaics\cite{10.1038/nmat2629} and quantum information\cite{PhysRevLett.97.053002,PhysRevLett.106.020501,doi:10.1021/nl201771h}. While surface plasmons rely on free electron oscillations, surface phonons exist because of the lattice vibrations in polar crystals.

Graphene has been shown to be good candidate for tunable plasmonics in the mid-IR and terahertz range\cite{doi:10.1021/nn406627u, 10.1038/nnano.2011.146, doi:10.1021/nl201771h, 10.1038/nphoton.2012.262, 10.1038/nphoton.2013.57, 10.1038/nature11254}, owing to the possibility of electrostatic doping\cite{10.1038/nature11253} and its ability to produce higher confinement and lower losses compared to metals\cite{PhysRevB.80.245435}. On the other hand, recently near-field imaging has shown that phonon polaritons in hexagonal boron nitide (hBN) possess extremely high confinement and even lower loss compared to graphene plasmon polaritons\cite{Dai07032014}. hBN shows natural hyperbolicity\cite{10.1038/nmat4149} which can potentially be used to explore exotic photonic properties\cite{10.1038/nphoton.2013.243} such as strong spontaneous emission enhancement\cite{:/content/aip/journal/apl/100/18/10.1063/1.4710548,10.1155/2012/452502}, negative refraction\cite{MOP:MOP10887} and thermal radiation enhancement\cite{:/content/aip/journal/apl/101/13/10.1063/1.4754616}. Since both graphene plasmons and hBN phonons reside in the mid-IR, the optical properties of graphene-hBN heterostructures would allow one to marry the advantage of their constituents, electrical tunability in the former and high quality factor of the latter, through their hybrid plasmon-phonon polaritons.

The study of the optical properties of graphene-hBN heterostructures are also motivated by the following recent developments. Firstly, hBN is now being used as a substrate of choice due to the preservation of high carrier mobility, as opposed to conventional SiO$_2$ substrates\cite{10.1038/nnano.2010.172}. The higher carrier mobility also translates to better plasmon quality factors\cite{10.1038/nmat4169}. Secondly, phonon modes of hBN can couple to graphene plasmon providing the possibility of observing interesting effects such as phonon-induced transparency\cite{Tony_hBN_G_paper, doi:10.1021/nl501628x}. Recently, a study of patterned graphene on monolayer hBN revealed a coherent coupling between plasmon modes in graphene and optical phonon modes in single layer hBN\cite{doi:10.1021/nl501096s}. As shown in \cite{Dai07032014}, hBN thin films can support several higher order phonon-polaritonic waveguide modes inside the Reststrahlen band. These modes show a dispersion which can be efficiently controlled by varying the thickness of the slab. In addition, both graphene and hBN can be grown in large area using CVD techniques\cite{doi:10.1021/nl1022139,doi:10.1021/nl801827v}, hence allowing in principle the construction of arbitrary heterostructures multilayers stack. The combination of these properties render graphene-hBN heterostructure an interesting photonic system.

  \begin{figure*}
  \begin{subfigure}[b]{3.8in}
  \centering
            \includegraphics[width=3.8in]{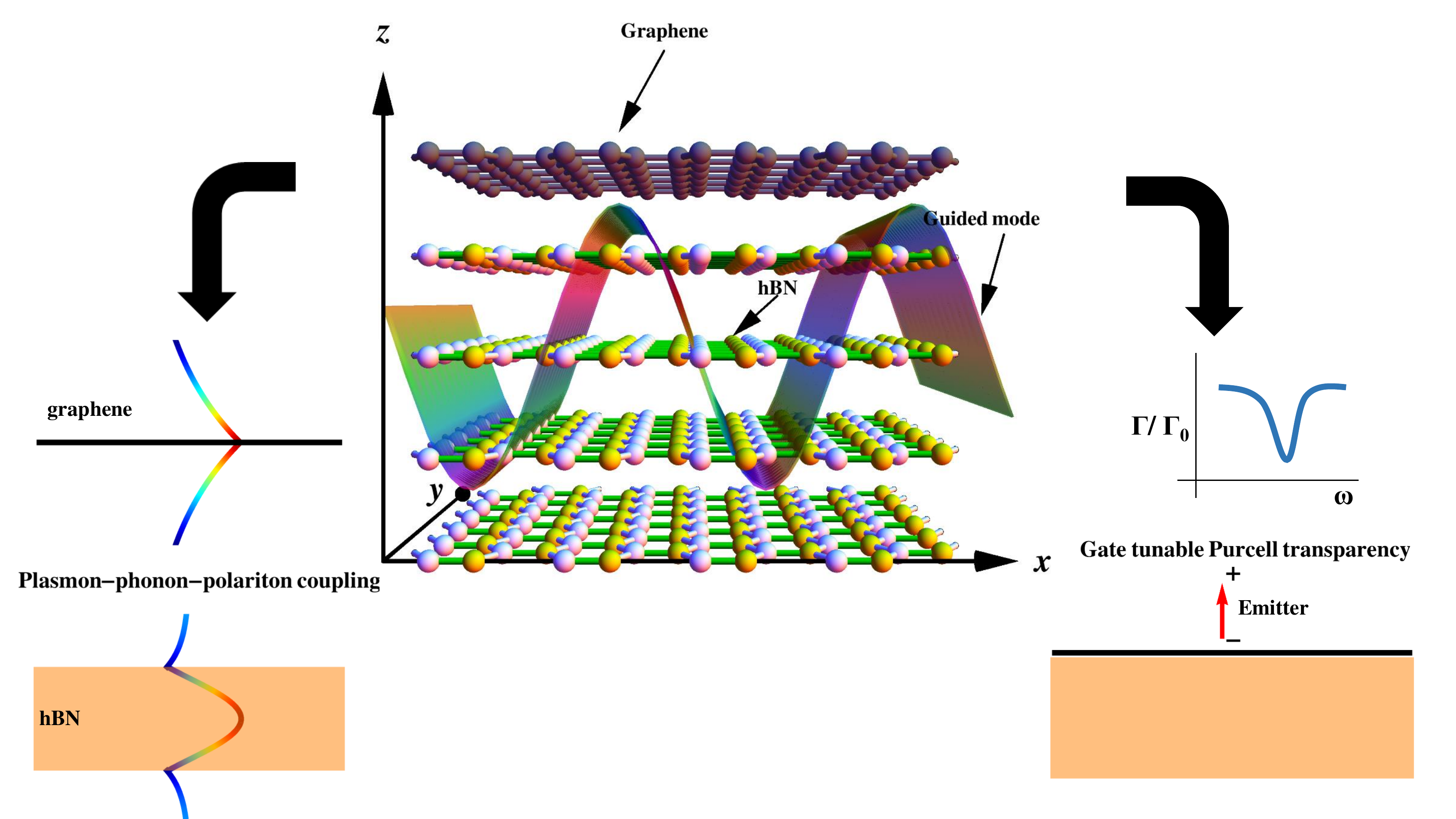}
  \caption{}
  \label{fig:hBN_G_schematic}
\end{subfigure}
\begin{subfigure}[b]{2.5in}
  \centering
  \includegraphics[width=2.5in]{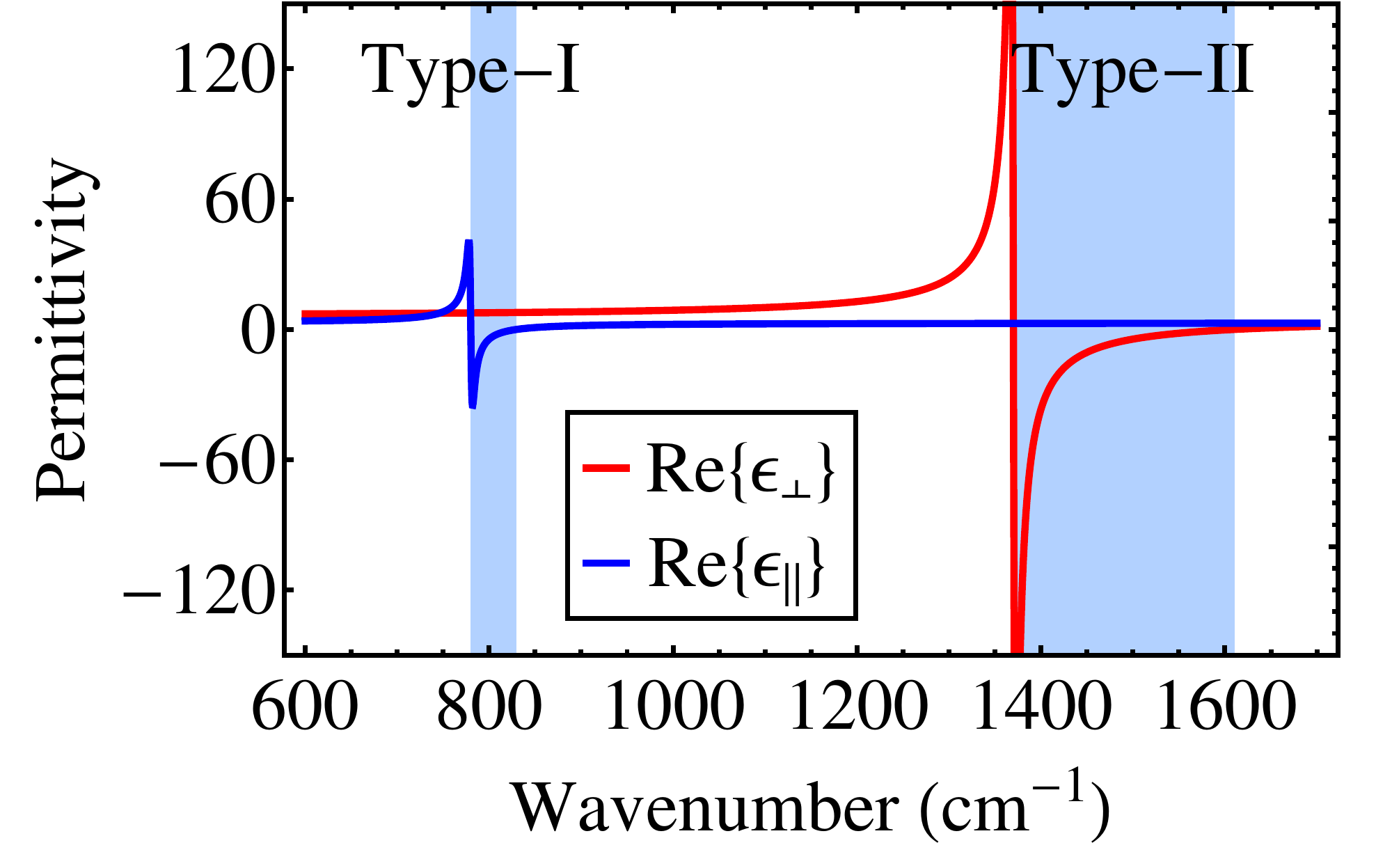}
  \caption{}
  \label{fig:eps_hBN}
\end{subfigure}
\caption{(a)\textit{Schematic of the geometry}: Monolayer graphene deposited on an hBN film of thickness $t_{hBN}$. The figure is not to scale. This work focuses on the interaction of graphene plasmon with phonon in hBN (left) and the emission of a nearby dipole into these hybrid modes (right). (b)\textit{Permittivity of hBN}: Permittivity tensor components of hBN clearly show the possibility of hyperbolicity in certain frequency ranges. The relevant parameters for the phonon frequencies were obtained from \cite{Cai2007262}.}
\end{figure*}

Typical configuration of the graphene-hBN heterostructure studied in this work, that is, monolayer graphene deposited on hBN thin film, is shown in Fig.~\ref{fig:hBN_G_schematic}. First we will describe the nature of the couplings between graphene plasmons and phonons of hBN for two different regimes of the hyperbolicity of hBN. Secondly, we exploit this coupling to demonstrate the possibility of inducing a dip in the Purcell spectrum due to this plasmon-phonon coupling and show its tunability via the external knobs of hBN slab thickness and of active tuning using electrostatic gating.

\section{Optical response of hBN and graphene}
\textit{Hexagonal boron nitride: }
Hyperbolic materials are anisotropic materials where the relative permittivity tensor is such that one of the three components has a sign different from the other two. This property leads to a hyperbolic or indefinite dispersion for electromagnetic waves propagating inside such a material, which results in exotic photonic properties\cite{10.1038/nphoton.2013.243}. Until recently, most practical realizations of hyperbolic media relied on artifically engineered systems, or the so called metamaterials. However, the recent discovery of a natural hyperbolicity in hexagonal boron nitride (hBN) crystals would make possible the design of atomic scale hyperbolic metamaterials, and might potentially allow one to cross over to regimes beyond the simple effective medium description in conventional hyperbolic metamaterials.

hBN is a van der Waals crystal with two kinds of IR active phonon modes relevant to hyperbolicity: 1) out of plane $A_{2u}$ phonon modes which have $\omega_{TO}=780 cm^{-1}$, $\omega_{LO}=830 cm^{-1}$ and 2) in-plane $E_{1u}$ phonon modes which have $\omega_{TO}=1370 cm^{-1}$, $\omega_{LO}=1610 cm^{-1}$\cite{Cai2007262}. This leads to two distinct Reststrahlen (RS) bands, where the lower frequency RS band corresponds to type-I hyperbolicity ($\epsilon_{\parallel}<0, \epsilon_{\perp}>0$) and the upper RS band shows type-II hyperbolicity ($\epsilon_{\perp}<0, \epsilon_{\parallel}>0$). The hBN permittivity is given by
\begin{equation}
\epsilon_m=\epsilon_{\infty,m}+\epsilon_{\infty,m}\cdot\frac{(\omega_{LO,m})^2-(\omega_{TO,m})^2}{(\omega_{TO,m})^2-\omega^2-\imath\omega\Gamma_m}
\label{eq:eps_hBN}
\end{equation}
where $m=\perp,\parallel$. The parameters employed in the above equation are taken from \cite{Cai2007262}. In addition to the LO and TO frequencies which have been mentioned in the previous paragraph, the other parameters are $\epsilon_{\infty,\perp}=4.87$, $\epsilon_{\infty,\parallel}=2.95$, $\Gamma_{\perp}=5$ cm$^{-1}$ and $\Gamma_{\parallel}=4$ cm$^{-1}$.

\textit{Graphene: }
The graphene response is modeled using local random phase approximation (local RPA). At temperature T, the two dimensional conductivity of graphene is given by \cite{10.1140/epjb/e2007-00142-3}:
\begin{flalign}
\sigma_{RPA}&(\omega) = \frac{2e^2 kT}{\pi\hbar^2}\frac{i}{\omega+i/\tau}\ln\left|2\cosh\left(\frac{\mu}{2kT}\right)\right| \nonumber \\
&+ \frac{e^2}{4\hbar}\left[H(\omega/2, T) + \frac{4i\omega}{\pi}\int_0^{\infty} d\zeta \frac{H(\zeta,T) - H(\omega/2, T)}{\omega^2-4\zeta^2} \right]
\label{eq:RPA}
\end{flalign}
where $H(\omega,T)=\sinh(\hbar\omega/kT)/[\cosh(\mu/kT)+\cosh(\hbar\omega/kT)]$. The first term in Eq. (\ref{eq:RPA}) represents intraband contribution and the remaining terms are contributions of the interband transitions to the total graphene conductivity. Here $\tau$ is the electron relaxation time. While Landau damping itself is already included in the conductivity model, the relaxation time typically has other contributions from 1) impurity scattering, 2) scattering with phonons ($\hbar\omega_{OPh}= 0.2$ eV) in graphene and phonon modes of polar substrates, 3) higher order processes such as phonon coupled to e-h excitations (which have to be treated separately), etc.\cite{10.1038/nphoton.2013.57, 10.1038/nmat4169, 10.1109/JPROC.2013.2260115}. In literature, relaxation times as long as 1000 fs have been reported  \cite{10.1038/nnano.2010.172, 10.1016/j.ssc.2008.02.024}. For frequencies larger than the optical phonon frequency of graphene, typically $\tau\sim 50$ fs  \cite{10.1038/nphoton.2013.57}. In this work, we use a graphene DC mobility $\mu_{DC}$ of $10,000$ cm$^2$V$^{-1}$s$^{-1}$. The mobility and relaxation time are related by $\tau=\mu_{DC}E_F/ev_F^2$\cite{PhysRevB.80.245435}.

\section{Plasmon-phonons: two regimes of mode coupling}

We consider the geometry as shown in Fig.~\ref{fig:hBN_G_schematic}. In this general case, by invoking the quasistatic approximation, the modal dispersion inside the two RS bands can be written as:
\begin{flalign}
q(\omega) = -\frac{\psi}{t_{hBN}} \biggl[ \tan^{-1}\left\{\frac{\epsilon_a+\imath (q/k_0)Z_0\sigma}{\epsilon_{\perp}\psi} \right\} + \tan^{-1}\left\{\frac{\epsilon_s}{\epsilon_{\perp}\psi} \right\} + \pi n\biggr]
 \label{eq:quasistatic_disp}
\end{flalign}
where $\psi=\sqrt{\epsilon_{\parallel}/\epsilon_{\perp}}/\imath$, $\sigma$ is the conductivity of the graphene, $t_{hBN}$ is the thickness of the hBN film, $\epsilon_a$ and $\epsilon_s$ are the relative permittivities of air and the substrate. On the other hand, outside the RS bands, the above condition is not sufficient since $\psi$ becomes imaginary. Thus, outside the two RS bands, the modal dispersion takes the form:
\begin{flalign}
q(\omega) = \frac{1}{2}\frac{\imath\psi}{t_{hBN}}\ln \biggl[ \left( \frac{1-\imath\psi (\epsilon_{\perp}/\epsilon_a)+\imath (q/k_0)(Z_0\sigma/\epsilon_a)}{1+\imath\psi (\epsilon_{\perp}/\epsilon_a)+\imath (q/k_0)(Z_0\sigma/\epsilon_a)} \right)\cdot  
 \left(\frac{1-\imath\psi (\epsilon_{\perp}/\epsilon_s)}{1+\imath\psi (\epsilon_{\perp}/\epsilon_s)} \right)\biggr]
 \label{eq:quasistatic_disp_outside}
\end{flalign}
In the lossless case, inside the two RS bands, $\psi$ is real whereas $\imath\psi>0$ outside the RS bands. It should be noted that for all subsequent calculations, we have used full-wave calculations, but we will refer to the quasistatic dispersion equations given above for intuitive understanding (see supplemental information for a comparison between quasistatic and full-wave results).

\begin{figure*}
\begin{subfigure}[b]{3.2in} 
\begin{framed}
\begin{subfigure}[b]{1.4in} 
  \includegraphics[width=1.4in]{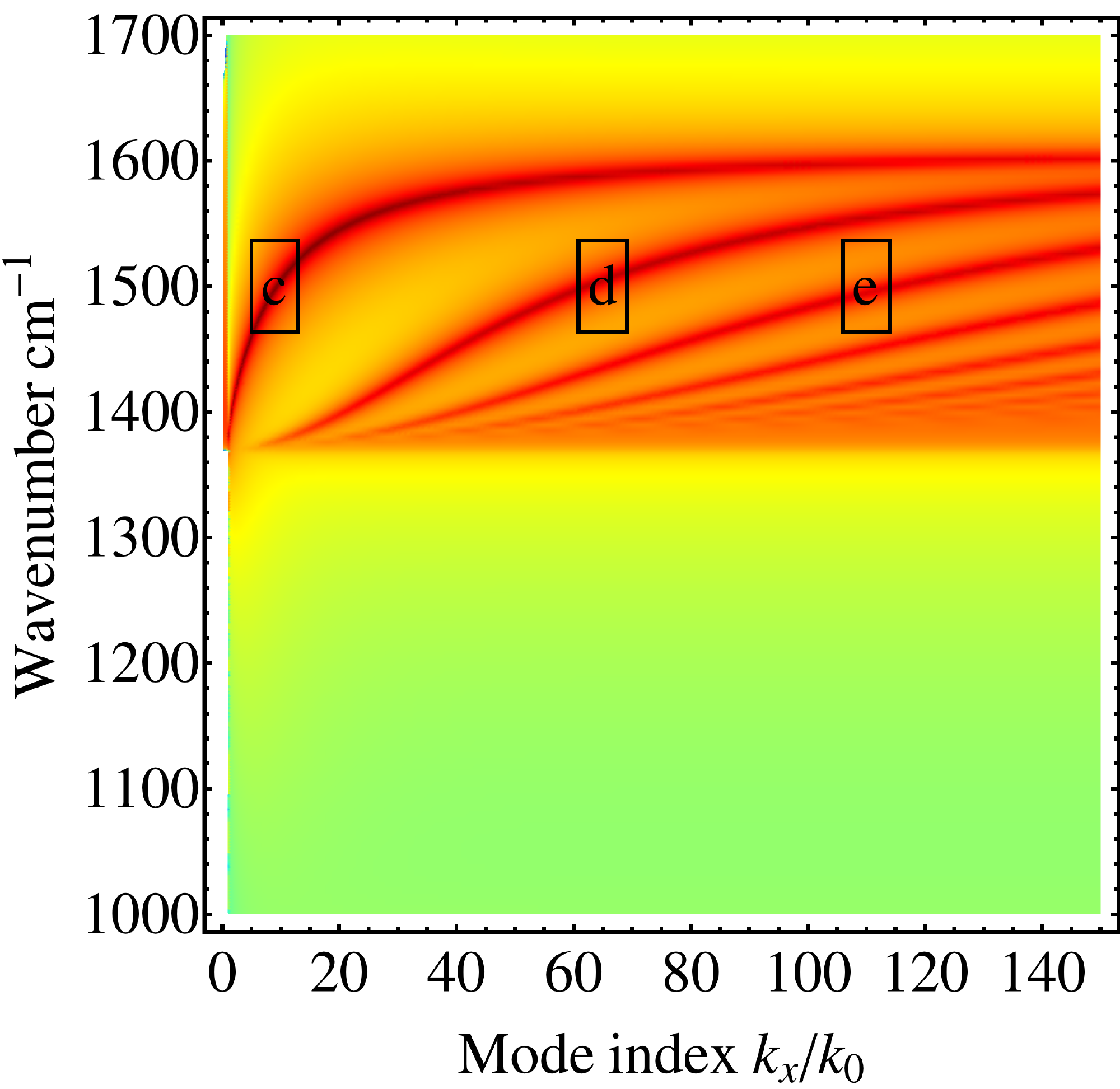}
  \caption{}
  \label{fig:AHA_disp_in_plane_50nm}
\end{subfigure}%
\begin{subfigure}[b]{1.7in} 
  \includegraphics[width=1.7in]{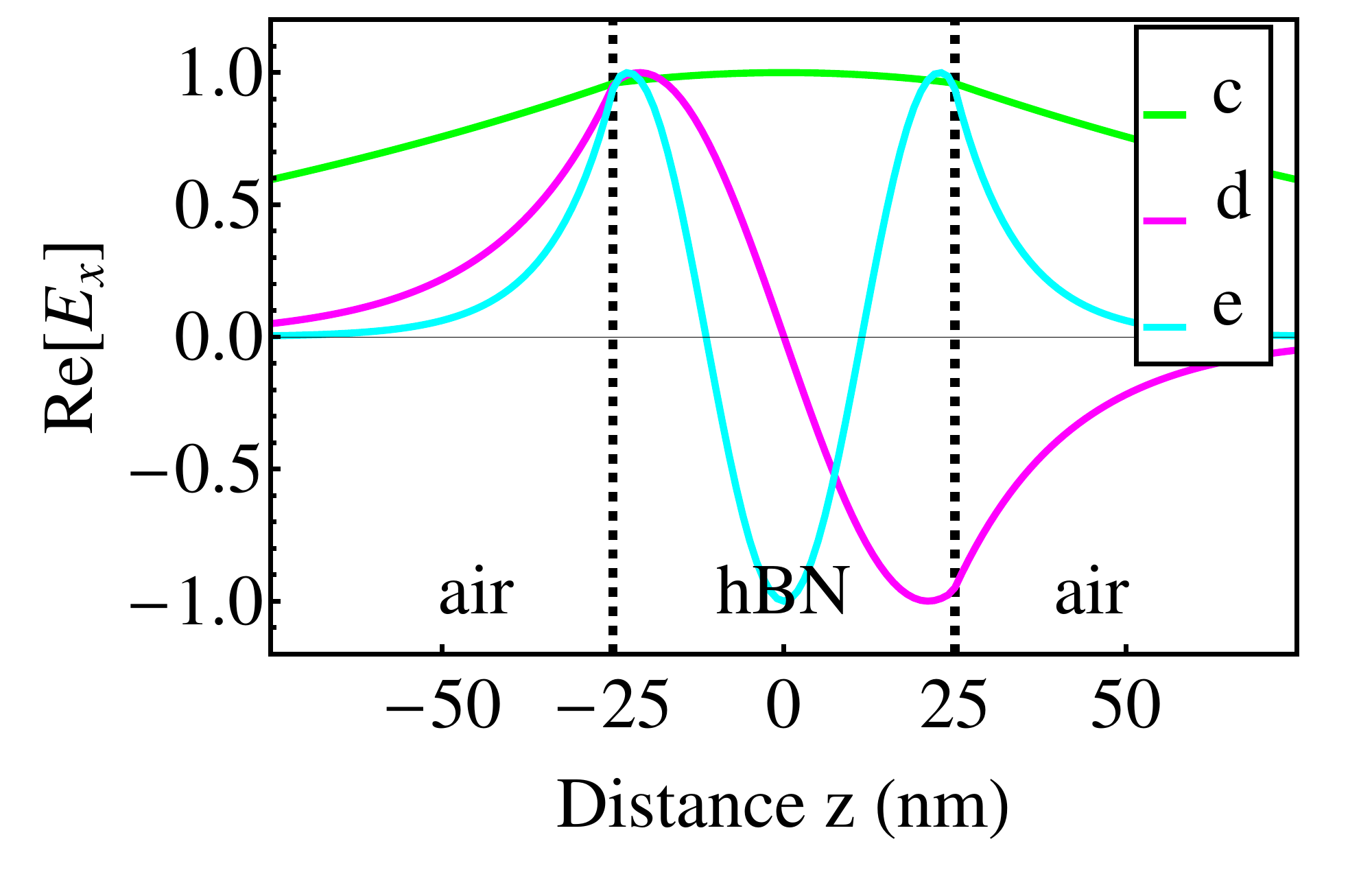}
  \caption{}
  \label{fig:AHA_Ex_in_plane_50nm}
\end{subfigure}\\

\vspace{0.5cm}
\begin{subfigure}[b]{1.4in} 
  \includegraphics[width=1.4in]{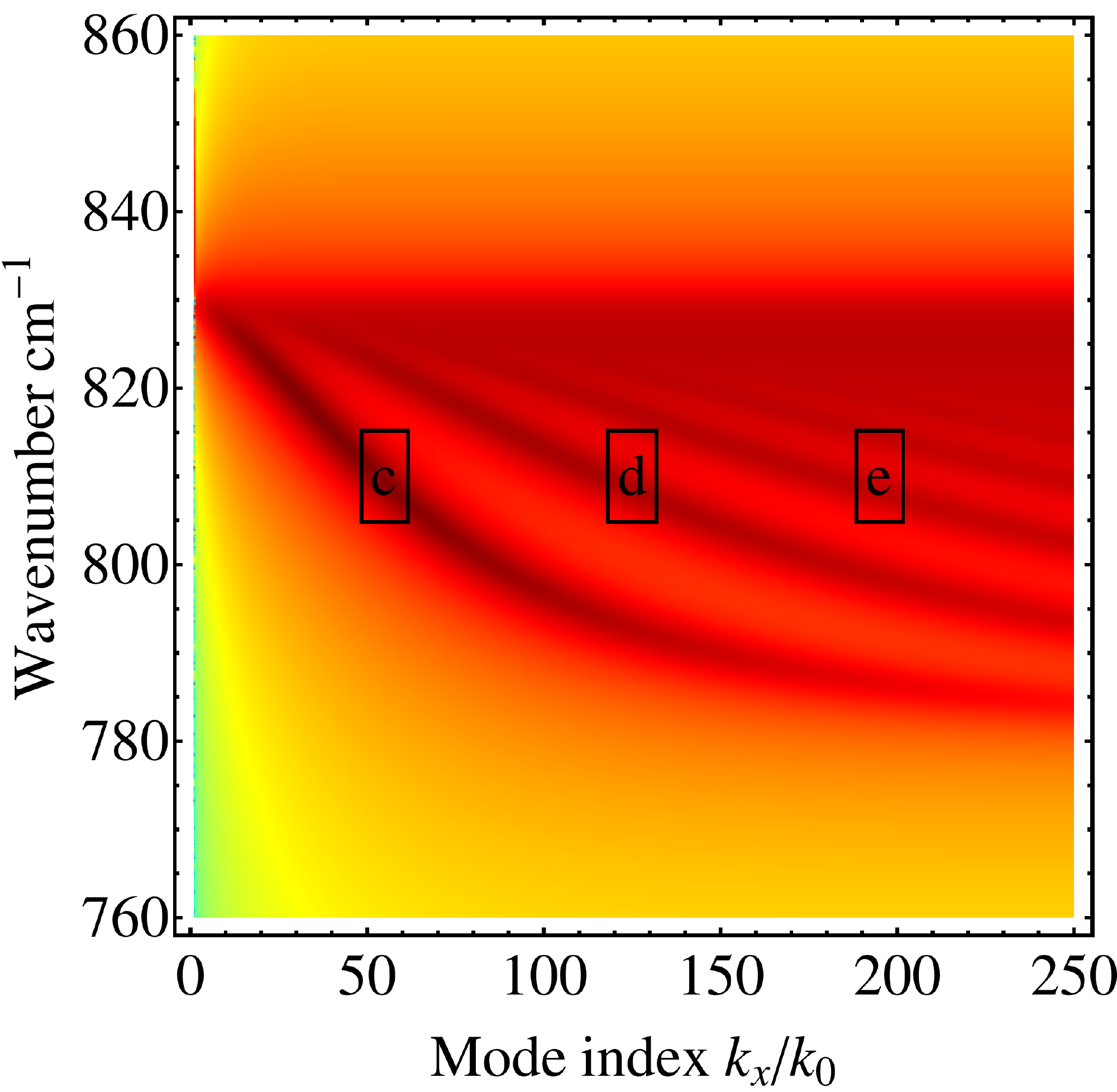}
  \caption{}
  \label{fig:AHA_disp_out_of_plane_50nm}
\end{subfigure}%
\begin{subfigure}[b]{1.7in} 
  \includegraphics[width=1.7in]{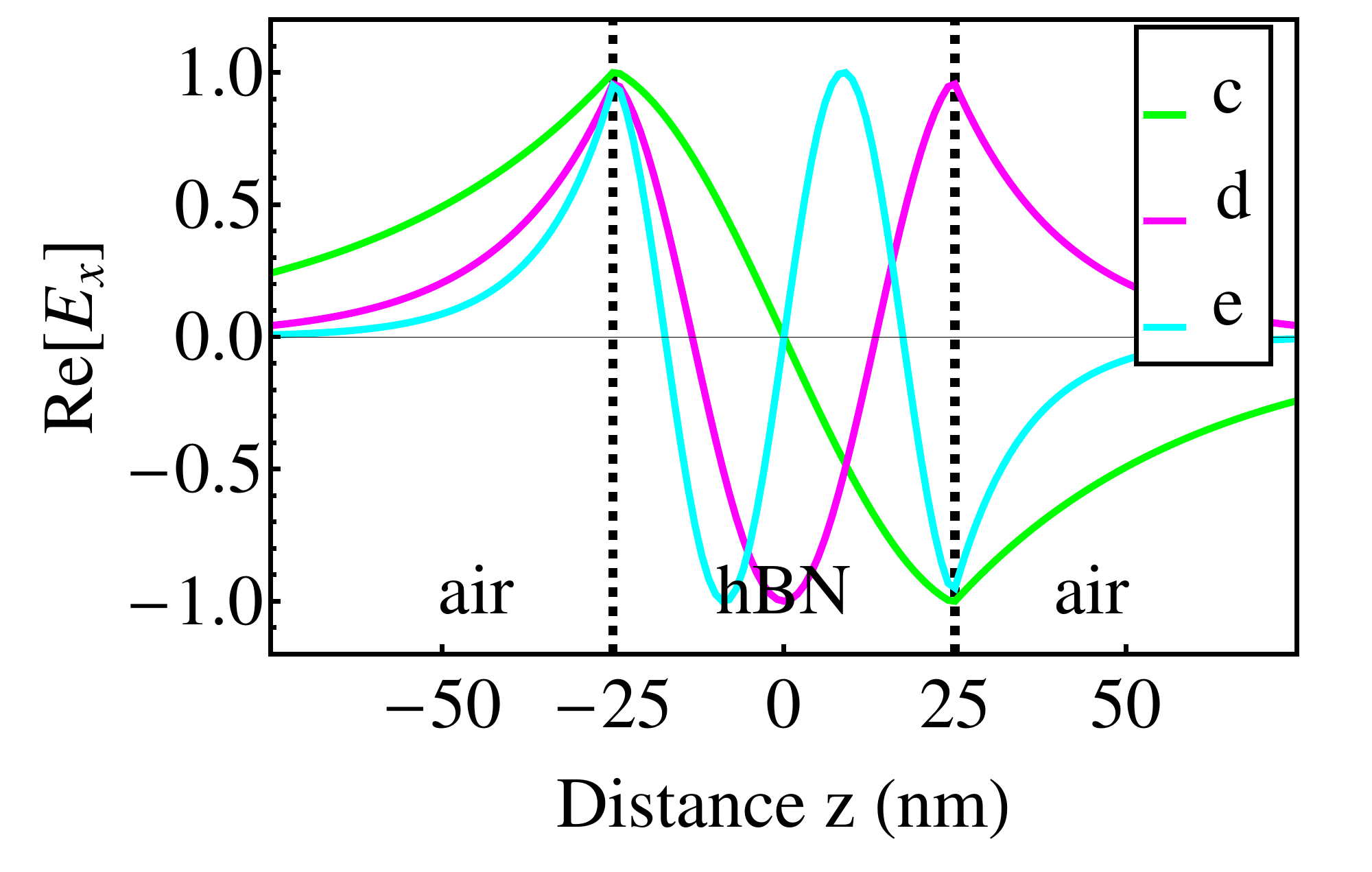}
  \caption{}
  \label{fig:AHA_Ex_out_of_plane_50nm}
\end{subfigure}\\

\caption*{hBN (AHA)}
\end{framed}
\end{subfigure}
\begin{subfigure}[b]{3.2in} 
\begin{framed}
\begin{subfigure}[b]{1.4in} 
  \includegraphics[width=1.4in]{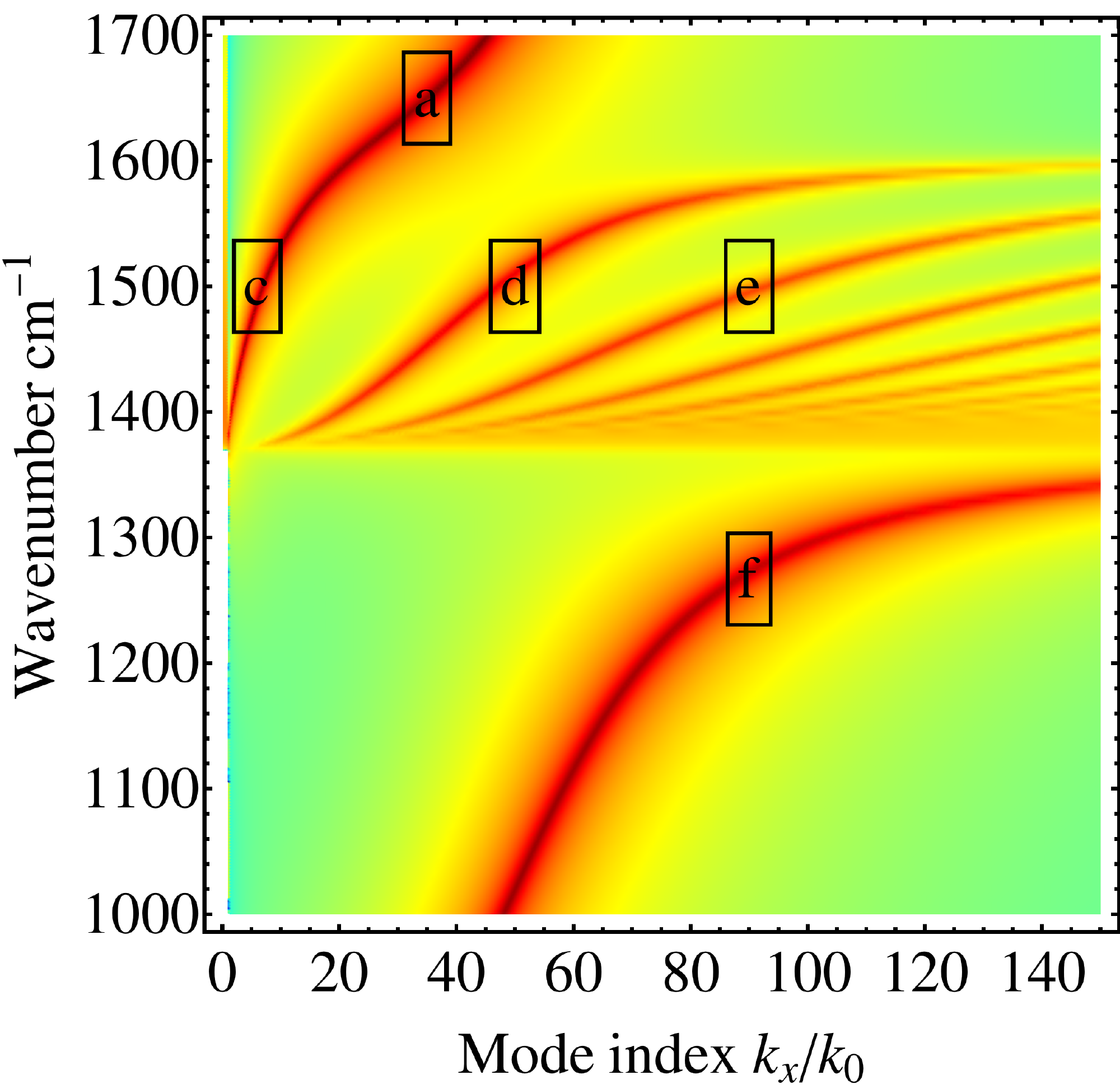}
  \caption{}
  \label{fig:AGHA_disp_in_plane_50nm}
\end{subfigure}%
\begin{subfigure}[b]{1.7in} 
  \includegraphics[width=1.7in]{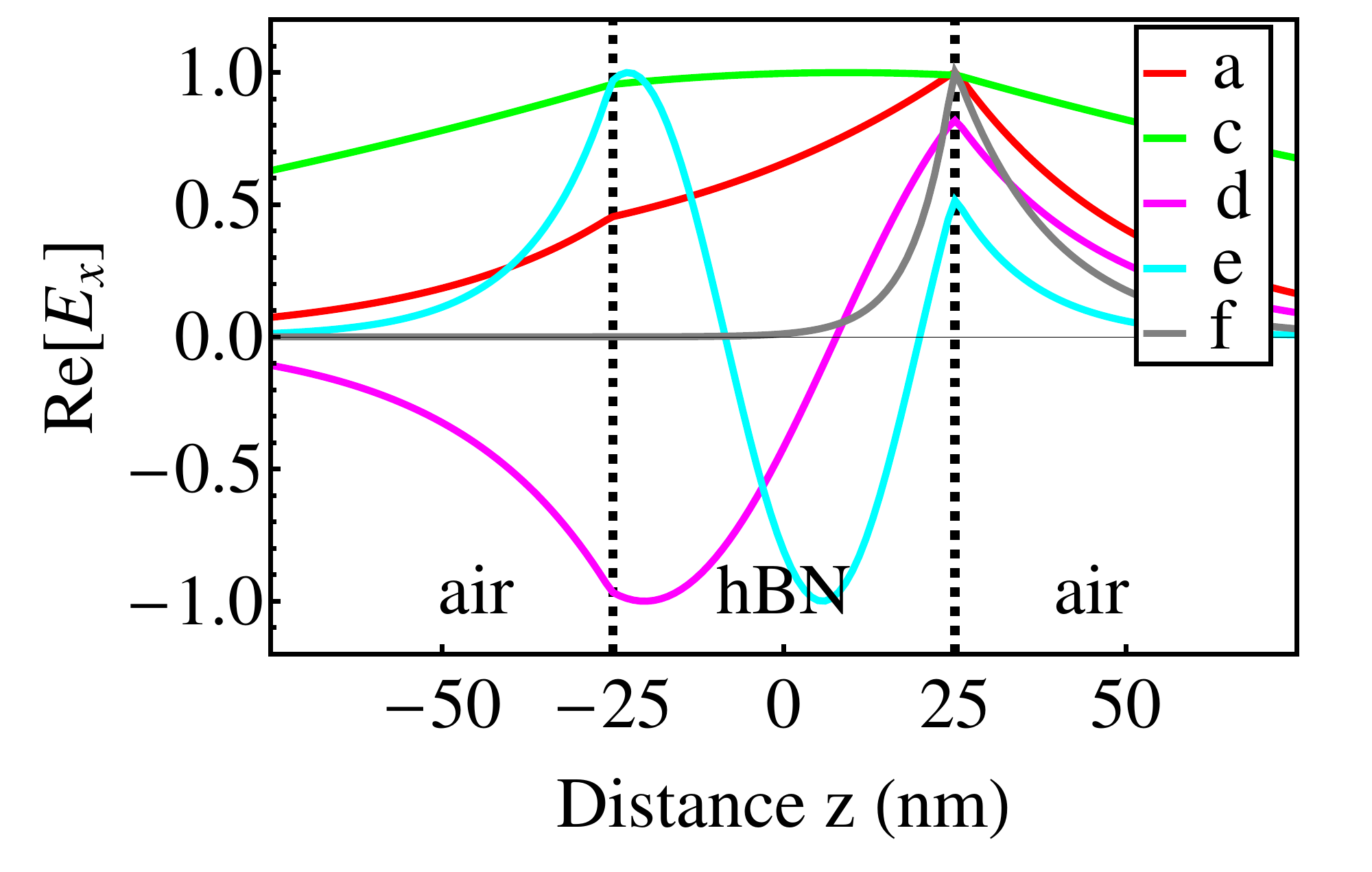}
  \caption{}
  \label{fig:AGHA_Ex_in_plane_50nm}
\end{subfigure}\\

\vspace{0.5cm}
\begin{subfigure}[b]{1.4in} 
  \includegraphics[width=1.4in]{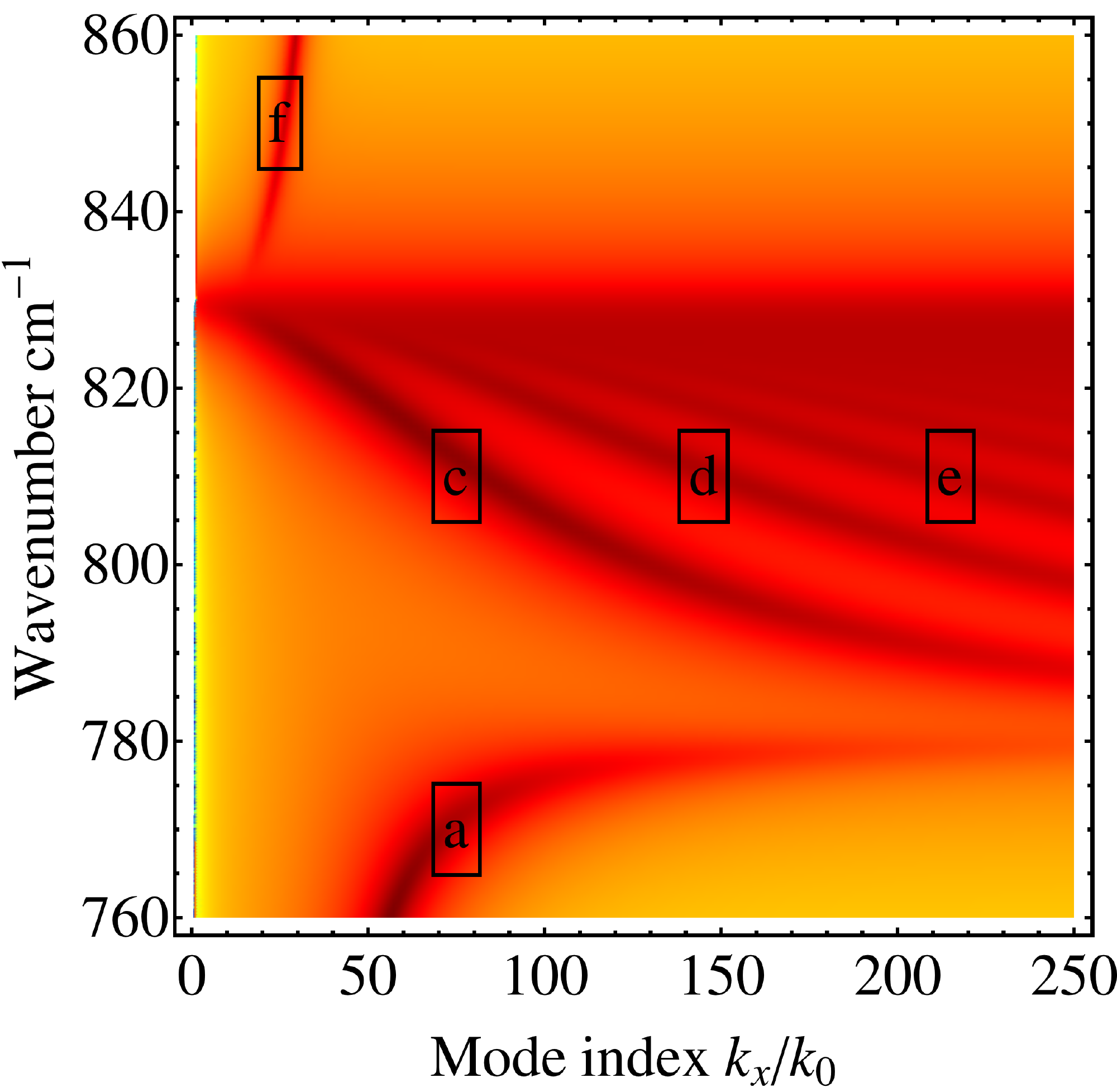}
  \caption{}
  \label{fig:AGHA_disp_out_of_plane_50nm}
\end{subfigure}%
\begin{subfigure}[b]{1.7in} 
  \includegraphics[width=1.7in]{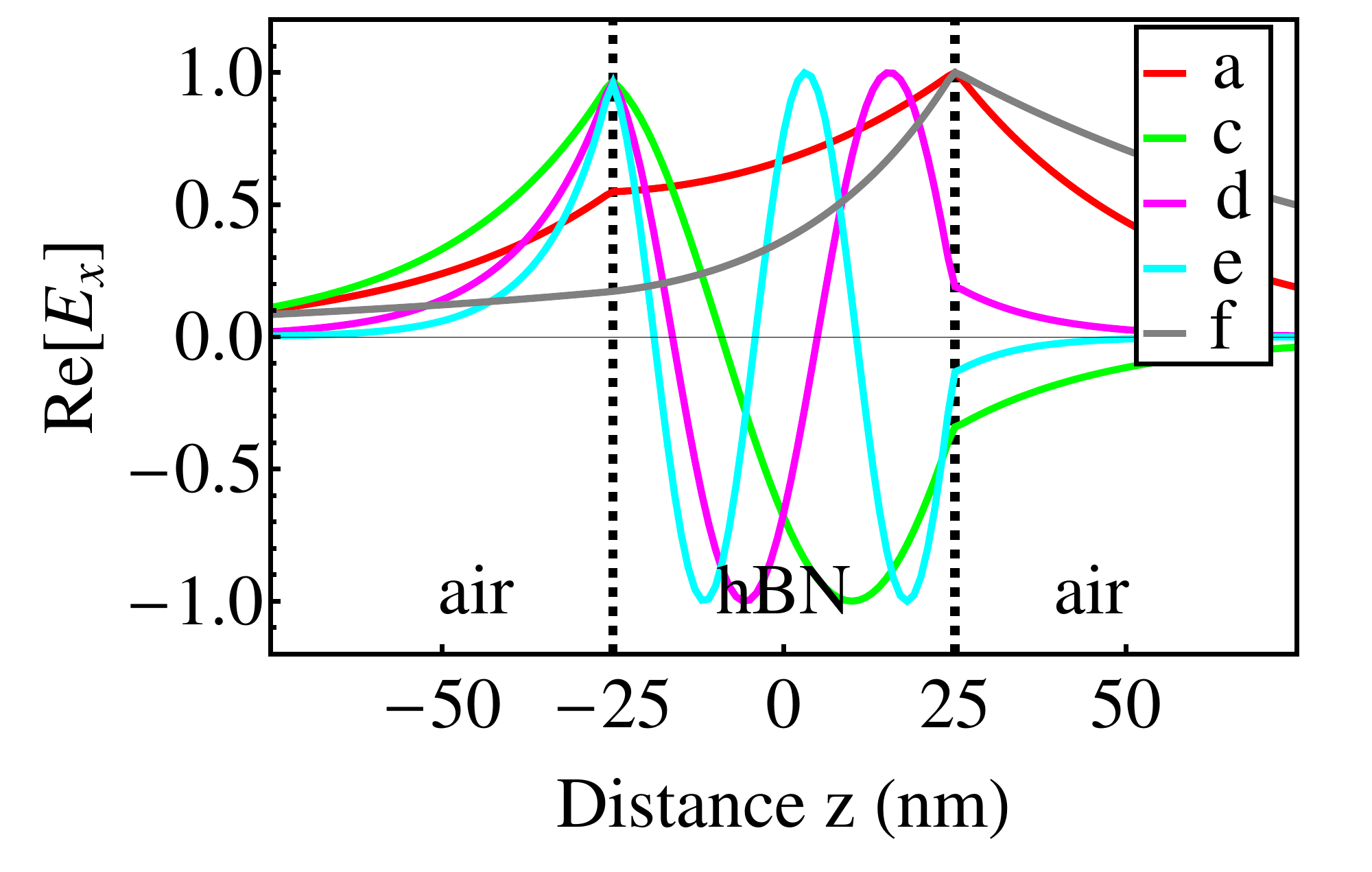}
  \caption{}
  \label{fig:AGHA_Ex_out_of_plane_50nm}
\end{subfigure}\\

\caption*{Graphene-hBN (AGHA)}
\end{framed}
\end{subfigure}

\caption{{\it Dispersion curves and mode profiles for phonon-polaritons in hBN thin slab (AHA) and coupled phonon-polariton plasmon modes in graphene-hBN system (AGHA): } Subfigures (a), (b), (e) and (f) represent the dispersion and mode profiles near the in-plane resonance; (c), (d), (g) and (h) represent those near the out-of-plane resonance of hBN. The alphabetical labels at different $(k_x,\omega)$ points on each of the dispersion diagrams are independent of each other and only correspond to the field profiles on their immediate right. Thickness of hBN is $t_{hBN}=\SI{50}{nm}$ and graphene Fermi level is assumed to be $\SI{0.5}{eV}$. Here AHA and AGHA denote air-hBN-air and air-graphene-hBN-air respectively.}
  \label{fig:fields_disp_hBN_in_plane_100nm}
\end{figure*}

To understand the plasmon-phonon coupling, we consider two cases: a) air-hBN-air (AHA) and b) air- graphene- hBN-air (AGHA). A third case of air- graphene- hBN- graphene- air (AGHGA) has also been presented in the supplementary information. We have chosen the geometry of symmetric waveguides (the two surrounding half-spaces are both assumed to be air) only for simpler presentation, but all our results and discussions in this section have been verified and hold true for asymmetric waveguides, which is the common situation experimentally. Later in this paper, when we do consider a substrate, these geometries will be denoted as AHS, AGHS and AGHGS instead. The refractive index of this dummy substrate was chosen to be 1.5. 

The phonon-plasmon-polariton dispersion is highlighted in the frequency range close to the two RS bands in Fig.~\ref{fig:AGHA_disp_in_plane_50nm} and \ref{fig:AGHA_disp_out_of_plane_50nm}. The nature of these modes is completely different between the in-plane and out-of-plane RS bands owing to differing types of hyperbolicity. This is described next two subsections. Different values of $k_x$ on these dispersion diagrams can be experimentally accessed for instance by fabricating finite width graphene nanoribbons on top of hBN\cite{10.1038/nphoton.2013.57}. All the dispersion curves are represented by a density plot of the imaginary part of the Fresnel reflection coefficient for p-polarization in $(\omega, k_x)$ space. The field profiles were obtained using \textsc{comsol multiphysics} package.

\subsection{Plasmon-phonons for the in-plane RS bands}

This is the frequency range which shows type-II hyperbolicity, which means that the in-plane dielectric function is negative and the out of plane component is positive. Therefore, for this band the quantity $\psi$ in Eq.~\ref{eq:quasistatic_disp} is less than zero also. For the AHA case, this gives several slab polariton modes for $n=0, 1,2,3,...$ as suggested in the dispersion in Fig.~\ref{fig:AHA_disp_in_plane_50nm}. Each of these mode numbers corresponds to the number of nodes in the tangential electric field $E_x$ as shown in Fig.~\ref{fig:AHA_Ex_in_plane_50nm}. It is apparent from Fig.~\ref{fig:AHA_Ex_in_plane_50nm} that the fields are mostly confined inside the hBN slab rather than the surrounding media or the interfaces. It should be noted that the modes with smaller number of nodes occur at higher frequencies. This is a consequence of type II hyperbolicity and the shape of the permittivity curve for the hBN slab.

When the top interface of the hBN slab is covered with graphene (AGHA case), we observe an interesting effect: the $n=0$ phonon polariton merges smoothly into a plasmon mode of the graphene as shown in Fig.~\ref{fig:AGHA_disp_in_plane_50nm}. Note that the phonon polariton modes inside the hBN are sinusoidal in the $z$ direction, whereas the plasmon mode is exponentially decaying. This is also apparent from a comparison of the field plot (Fig.~\ref{fig:AGHA_Ex_in_plane_50nm}) for frequency points labeled ``a" and ``c" between the AHA and the AGHA geometries where in the latter case, the field profile inside the hBN is deviating markedly from a sinusoidal profile and tending toward an exponentially decaying one. This has the additional effect of shifting the peak of the field from the bulk of hBN to the graphene.

{This smooth transition is unexpected especially given the fact that at the top of this RS band, the hBN undergoes a topological transition\cite{Krishnamoorthy13042012} from hyperbolic to elliptical dispersion. However, an abrupt change at such a transition is expected only in bulk crystal of hBN that is thick enough. Since we have a thin layer of hBN, this transition can be smoothened. This behaviour can also be understood in terms of a mode coupling picture. The graphene plasmon dispersion below this RS band starts becoming phonon-like as it approaches the TO frequency of the hBN. On the other hand, the lowest (zero) order phonon-polariton mode acquires a plasmonic character near the LO frequency. Higher order phonon polariton modes are not affected by this coupling due to symmetry mismatch which can be seen by looking at the field profiles of the modes ``d" and ``e" in Fig.~\ref{fig:AGHA_Ex_in_plane_50nm}. The plasmon mode ``f" does not show any node inside the hBN whereas all the phonon-polariton modes other than ``c" show one node or more. Thus mode ``c" which is the lowest order mode can transform smoothly into the plasmon. This hand waving argument can be rigorously substantiated by calculating the modal overlaps.}

In order to confirm this picture, we investigated whether by coupling to the acoustic plasmon mode of the double layer graphene we can make the first order phonon-polariton mode ``d" also acquire plasmonic character. This case (AGHGA) is presented in the supplementary information, where it is clearly seen that the first order phonon polariton also merges smoothly into the acoustic plasmon mode. This behaviour can be easily understood by noticing that the graphene plasmon has split into symmetric and antisymmetric modes with zero and one nodes which will couple only to the zeroth and first order phonon polariton modes respectively.

\subsection{Plasmon-phonons for the out of plane RS bands}

Phonon polariton modes in the out of plane RS bands for thick hBN slabs have received relatively less attention in literature so far since the out of plane phonon polariton response is relatively weak\cite{arXiv:1501.02343}. This band displays type I hyperbolicity where the out of plane dielectric function is negative instead of the in-plane one. This difference makes its coupling to the graphene plasmon markedly different from the in-plane phonon case discussed above.

For this band, $\psi$ in Eq.~\ref{eq:quasistatic_disp} is positive. Thus, in order to make the real part of the wavevector $q$ positive, we need $n < 0$. This is also reflected in the field plots in Fig.~\ref{fig:AHA_Ex_out_of_plane_50nm} where it is observed that there is no mode with zero number of nodes. In other words, the minimum value of $|n|$ is 1 instead of zero. A simple explanation of this behaviour can be obtained by reference to the relative signs of the Poynting vector and the wave-vector\cite{Huang:13}. In the case of out of plane resonance, since the dispersion is type-I, the Poynting vector component $S_z$ and the wavevector component $k_z$ point in the same direction. Thus, the dispersion equation $k_z t_{hBN} + \phi_r = n\pi$ does not admit $n=0$ solution since the left hand side is strictly positive. On the other hand, for the in-plane mode, type-II dispersion causes $S_z$ and $k_z$ to point in opposite directions which permits the left side approach zero value, thus allowing $n=0$ solution. Such behaviour also consistent with the predictions in \cite{Huang:13}.

In this case one can observe that the ordering of the modes in the region with negative group velocity is such that that the lower order mode occurs at smaller frequency, unlike the case of in-plane RS band. Again, this is due to the difference in the type of hyperbolicity.

In the AGHA case, it is observed that below the RS band, the plasmon mode becomes phonon like as it approaches the TO phonon frequency. The plasmon branch on the top of the RS band acquires a phonon-polariton character near $\omega_{LO}$. We also observe the effect of mode repulsion which causes the first order phonon polariton to blueshift compared to the case without graphene. As shown in the supplementary, this behaviour persists for the AGHGA case where both the graphene plasmon modes acquire phonon-polariton like character near $\omega_{LO}$ and $\omega_{TO}$. 

In summary, we observe that the coupling between the graphene plasmon and the hBN phonon-polaritons is completely different between the two RS bands. This is attributable to the different types of hyperbolicity associated with the in-plane and the out-of-plane phonon modes. In the next section, we will attempt to use this coupling for tuning the local density of states.

\section{Tunable spectral dips in emission}

When an excited quantum emitter is placed near a material system which can support a photonic mode, its lifetime is modified compared to the case when it emits in free space. This phenomenon is called Purcell effect\cite{PhysRev.69.674.2}. The emitter can typically be an excited atom, molecule or quantum dot.

The emitter can release its excitation energy into free space radiative modes as well as resonant modes and non-radiative or lossy modes of the material. The radiative contribution is modified due to the change of the boundary conditions for the electromagnetic fields because of the presence of the material. This results in the modification of far-field emission patterns\cite{Ford1984195}.

More interestingly, the emitter can also release its energy through the available resonant modes in the neighbouring matter. The strength of this light-matter interaction is governed by the ratio of quality factor and the volume of the available photonic mode. The basic idea of nanophotonics is to provide sub-wavelength mode volumes which enhance this interaction. For instance, the power of graphene plasmonics is in providing very small mode volumes\cite{doi:10.1021/nl201771h}. On the other hand, hBN phonon polaritons provide in addition, high quality factors as well\cite{Dai07032014}.

One key aspect of hyperbolic materials is that when moving from a hyperbolic dispersion regime to an elliptical one, a sharp change in the local density of states (LDOS) is observed\cite{Krishnamoorthy13042012}. This is due to the availability of high-k states in the hyperbolic band. To explore this phenomenon in the context of graphene-hBN slab system, we considered the spontaneous emission enhancement (Purcell effect) of a quantum emitter with polarization perpendicular ($\hat{z}$) to the hBN surface. This orientation of the emitter was chosen since the $\hat{z}$-dipole moment will only couple to the TM modes. Parallel polarization can also be treated in a similar manner. In the present case, the spontaneous emission rate of the $\hat{z}$-polarized dipole, also called the partial LDOS or PLDOS, is given by\cite{Krishnamoorthy13042012,2040-8986-14-6-063001}:
\begin{equation}
\frac{\Gamma}{\Gamma_0} = 1 + \frac{3}{2k_0^3}\int_{0}^{\infty} dk_x k_x \Re\left\{ \frac{k_x^2 e^{2\imath k_z d_s}r_p(\omega,k_x)}{k_z}\right\}
\label{eq:Purcell}
\end{equation}
where $\Gamma_0$ is the free space radiative decay rate, $d_s$ is the distance of the quantum emitter (source) from the hBN slab and $r_p(\omega,k_x)$ is the Fresnel reflection coefficient from the graphene coated hBN slab. This expression includes both radiative as well as non-radiative contributions to the total decay rate. The Purcell enhancement in the hyperbolic regime is clearly observable in the AHS case shown in Fig.~\ref{fig:Purcell_ds_dependence_out_of_plane_AHS} and \ref{fig:Purcell_ds_dependence_in_plane_AHS}, where the PLDOS rises sharply inside both the RS bands. As shown in the following, in the presence of graphene, a dip in the Purcell enhancement is observed, whose spectral width and location is strongly tunable both actively using electrostatic doping and also by changing the thickness of hBN. In particular, for the out-of-plane RS band, the reduction in the decay rate is consistently found to be about an order of magnitude in the presence of phonon-plasmon coupling. {The observed dips in Purcell spectrum arise are due to coupling between spectrally broad plasmon and much narrower phonon resonance. Such dips is analogous to induced transparency in the absorption spectrum observed recently in such systems\cite{PhysRevLett.112.116801}.} The induced transparency can be understood in the classical coupled oscillator picture with oscillators of contrasting damping rate\cite{doi:10.1021/nl501628x,PhysRevLett.112.116801}. Such a large modification in the PLDOS suggests possible application in tuning the radiative energy transfer to hBN via phonon modes. Note that the presence of graphene is also expected to enhance the nonradiative decay rate of the emitter via energy transfer into the plasmon mode\cite{doi:10.1021/nl201771h}.

\subsection{Purcell spectra versus emitter position}

\begin{figure*}
\centering
\begin{subfigure}{3.2in}
  \centering
  \includegraphics[width=3.2in]{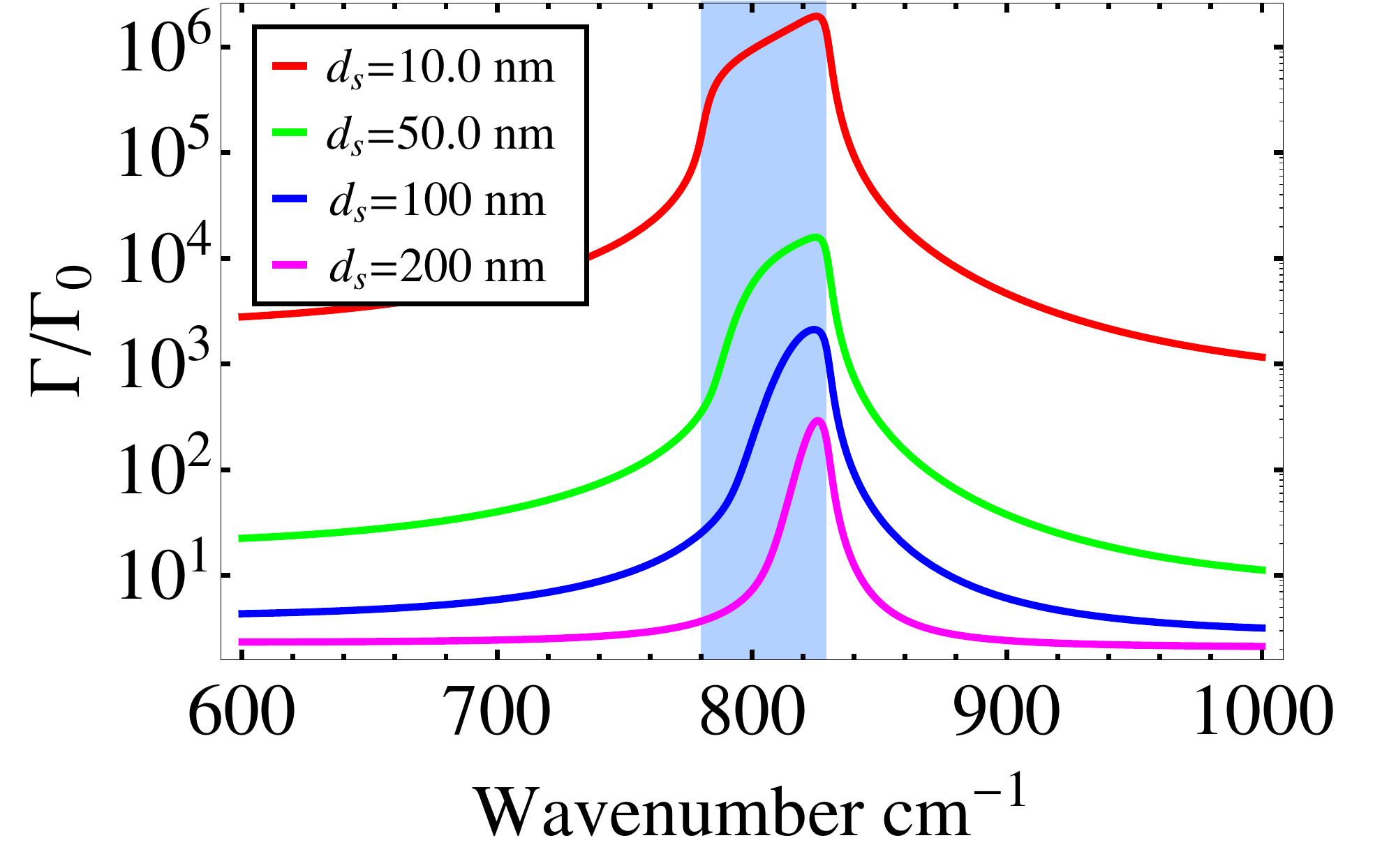}
  \caption{AHS}
  \label{fig:Purcell_ds_dependence_out_of_plane_AHS}
\end{subfigure} 
\begin{subfigure}{3in}
  \centering
  \includegraphics[width=3in]{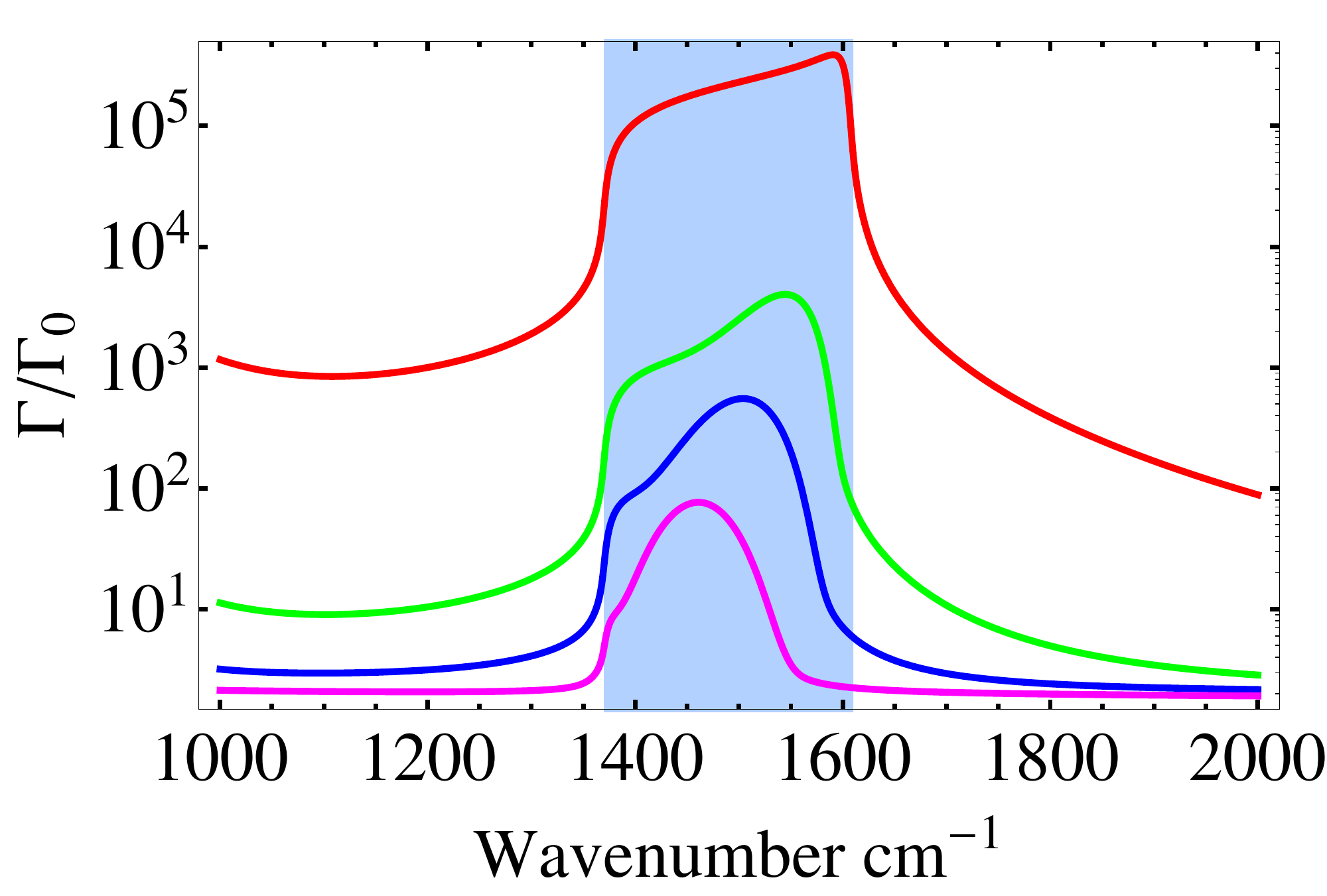}
  \caption{AHS}
  \label{fig:Purcell_ds_dependence_in_plane_AHS}
\end{subfigure}
\\
\begin{subfigure}{3in}
  \centering
  \includegraphics[width=3in]{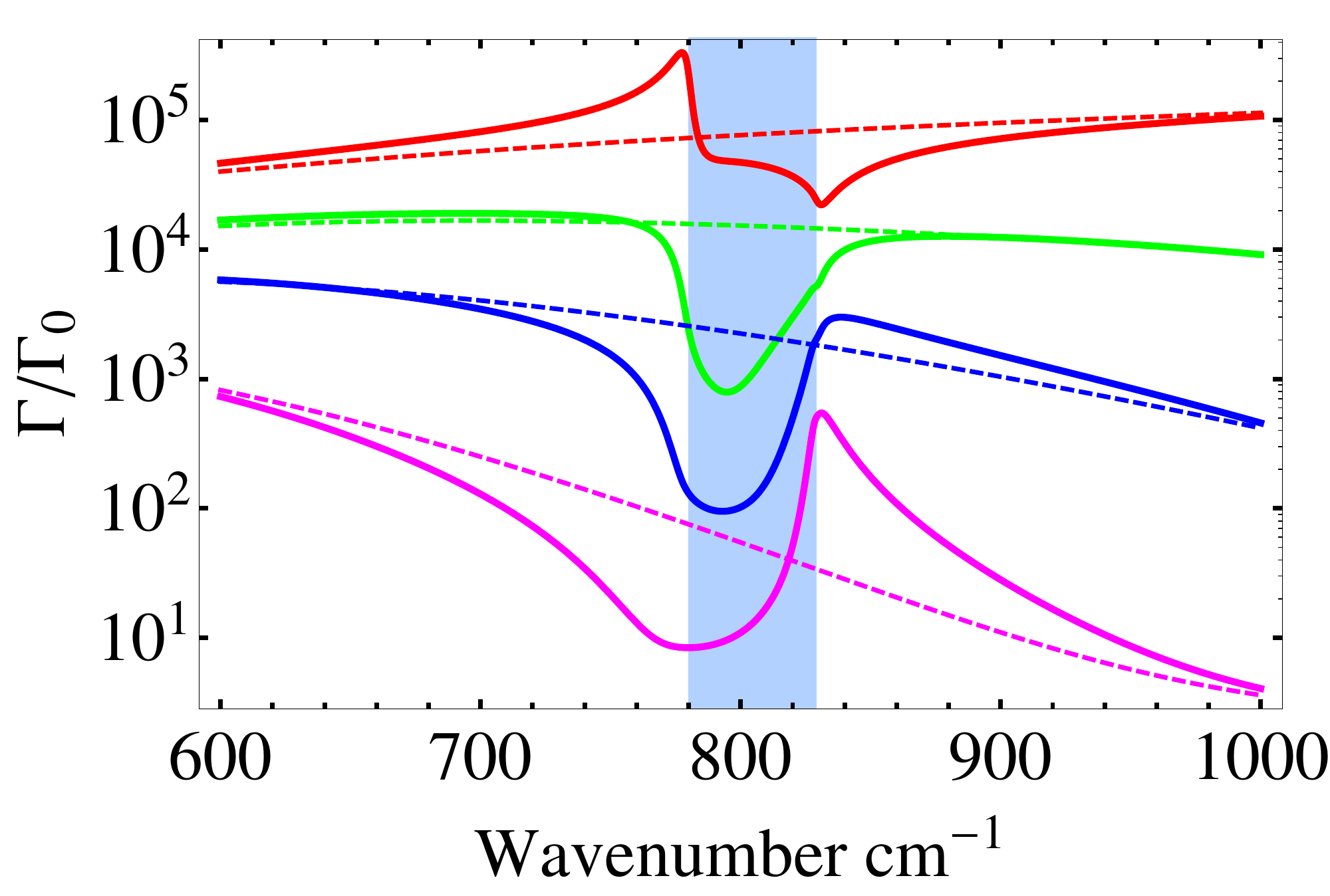}
  \caption{AGHS}
  \label{fig:Purcell_ds_dependence_out_of_plane_AGHS}
\end{subfigure} 
\begin{subfigure}{3in}
  \centering
  \includegraphics[width=3in]{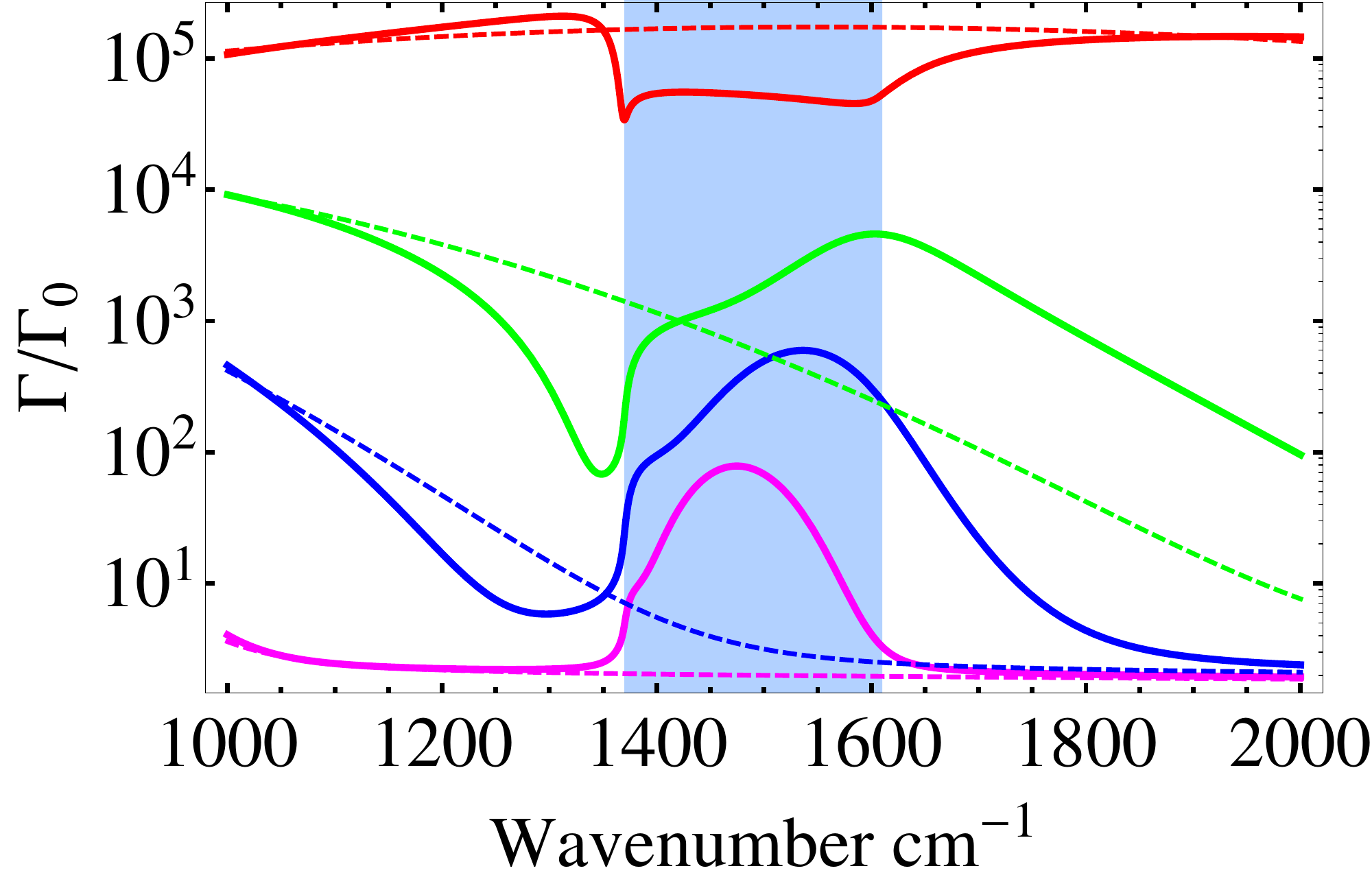}
  \caption{AGHS}
  \label{fig:Purcell_ds_dependence_in_plane_AGHS}
\end{subfigure} 
\caption{{\it Dipole distance dependent Purcell spectra in graphene-hBN system: } Distance $d_s$ of the quantum emitter from the hBN is varied. Subfigures (a) and (b) represent the spectra for the case of hBN thin film (AHS) for the two RS bands respectively. Subfigures (c) and (d) represent the case of monolayer graphene deposited on the hBN thin film (AGHS). Dashed lines show the Purcell enhancement due to graphene plasmon without the phonon-polariton contribution from the hBN slab, that is, assuming $\epsilon_{hBN}(\omega\rightarrow 0)$.  Thickness of hBN slab is fixed at $t_{hBN}=\SI{50}{nm}$. Graphene Fermi level is assumed to be $\SI{0.5}{eV}$. In all the figures, the emitter is on the air side.}
\label{fig:Purcell_ds_dependence}
\end{figure*}

The shape of the Purcell spectra in this case strongly depends on the dipole distance as can be seen in Fig.~\ref{fig:Purcell_ds_dependence_out_of_plane_AHS} and \ref{fig:Purcell_ds_dependence_in_plane_AHS}. To understand this variation we note that since we are in the near field, the integrand in Eq.~\ref{eq:Purcell} can be simplified as $\propto k_x^2 e^{-2k_x d_s}\Im\{r_p(\omega,k_x)\}$. As noted in \cite{PhysRevB.87.085421}, this form of the integrand leads to the maximum contribution to the PLDOS coming from $k_x\approx1/d_s$. This effective wave-vector cutoff together with the dispersion curve can be used to predict the shape of the Purcell spectra at different distances of the dipole emitter. For instance, if we look at the spectra near the out-of-plane modes in Fig.~\ref{fig:Purcell_ds_dependence_out_of_plane_AHS}, it is seen that as the dipole moves away from the hBN, the PLDOS peak shifts towards the upper part of the RS band, that is, towards $\omega_{LO}$. To understand this, we note from the dispersion curve shown in Fig.~\ref{fig:AHA_disp_out_of_plane_50nm} that at high $k_x$, the dispersion of the phonon-polariton modes becomes flatter, implying a high density of states available at the respective temporal frequencies. Thus at small $d_s$, all of these states are contributing to the PLDOS which remains high throughout the RS band. As the emitter is taken farther away from the the hBN, the maximum contributing $k_x$ moves from higher to lower values. At smaller values of $k_x$, the lower phonon-polariton branches start acquiring curvature which results in reduction of density states. Thus at these values of $k_x$ (or equivalently, $d_s$) only the high frequency phonon-polaritons which have a flatter dispersion, can contribute effectively to the PLDOS. This results the in PLDOS peak becoming sharper and moving upward in frequency as the dipole moves away from the hBN. A similar behaviour is observed for the in-plane phonon RS band in Fig.~\ref{fig:Purcell_ds_dependence_in_plane_AHS}, where due to the opposite shape of the dispersion curve, the PLDOS peaks move downward in frequency as the dipole separation from hBN increases.

When graphene is coated on top of this hBN slab, even more interesting spectral shapes are obtained. Most notably, due to the plasmon-phonon coupling, spectral dips of an order of magnitude can be obtained in the PLDOS for a range of dipole distances. Let us first focus on the out-of-plane RS band for the AGHS case, as shown in Fig.~\ref{fig:Purcell_ds_dependence_out_of_plane_AGHS}. It is clear from the dispersion in Fig.~\ref{fig:AGHA_disp_out_of_plane_50nm} that at large $k_x$, the plasmon dispersion below $\omega_{TO}$ is flat and this results in a peak in the PLDOS for smaller $d_s$ in Fig.~\ref{fig:Purcell_ds_dependence_out_of_plane_AGHS}. This plasmon peak peak seems to disappear as we move to larger $d_s$. This can be explained based on two observations. Firstly, since the plasmon dispersion acquires curvature at low $k_x$, there is a broadening of the plasmon peak since it is no longer flat. Secondly, at low $k_x$, the plasmon branch in the dispersion in Fig.~\ref{fig:AGHA_disp_out_of_plane_50nm} redshifts. Both these effects lead to an apparent disappearance of the PLDOS peak near $\omega_{TO}$ with increasing $d_s$. 

Next, let us try to understand what happens when we enter the RS band from below $\omega_{TO}$. If we are at sufficiently large $d_s$ (or low $k_x$), the plasmon is far below $\omega_{TO}$ and it has curvature, implying that it will contribute a broad peak located far below $\omega_{TO}$. Inside the RS band on the other hand, we know from previous analysis that there are flatter states available. Thus these two effects give the impression of a kind of ``band gap" being created near $\omega_{TO}$, which will result in a decrease in the PLDOS in between these two peaks since there are no states available. Note that the spectral width and location of this dip will depend on the value of $d_s$ that governs which $k_x$ contributes maximally. As shown in the supplementary, we can even make this ``dip" go away depending on the dipole distance $d_s$. The width of this dip is smallest at large $k_x$ (small $d_s$), which is explained by the changing curvature of the phonon-polariton branches as a function of $k_x$, as explained in the AHS case before. This explanation is consistent with our findings in Fig.~\ref{fig:Purcell_ds_dependence_out_of_plane_AGHS}.

Beyond $\omega_{LO}$, we encounter the plasmon mode again. Unlike the plasmon mode below $\omega_{TO}$, this plasmon mode is flatter at low $k_x$ because it acquires phononic character there. Thus the plasmon peak above the RS band is narrowest for large $d_s$. This explains the trend in the PLDOS peak arising above $\omega_{LO}$ in Fig.~\ref{fig:Purcell_ds_dependence_out_of_plane_AGHS}.

This spectral signature is carried over in a qualitatively similar way to the in-plane RS band as shown in Fig.~\ref{fig:Purcell_ds_dependence_in_plane_AGHS} with notable differences in the trend peak locations on account of the different type of hyperbolicity of this band. For instance, the dip below $\omega_{TO}$ in this case is always outside the RS band. This is because the higher order phonon-polariton modes for the in-plane branch cluster near $\omega_{TO}$ instead of $\omega_{LO}$ like in the out of plane case. This results in there being always states available at frequencies even slightly above $\omega_{TO}$.

\subsection{Purcell spectra versus hBN thickness}
\begin{figure*}
\centering
\begin{subfigure}{3.2in}
\begin{subfigure}[b]{2.2in}
\centering
\includegraphics[width=2.2in]{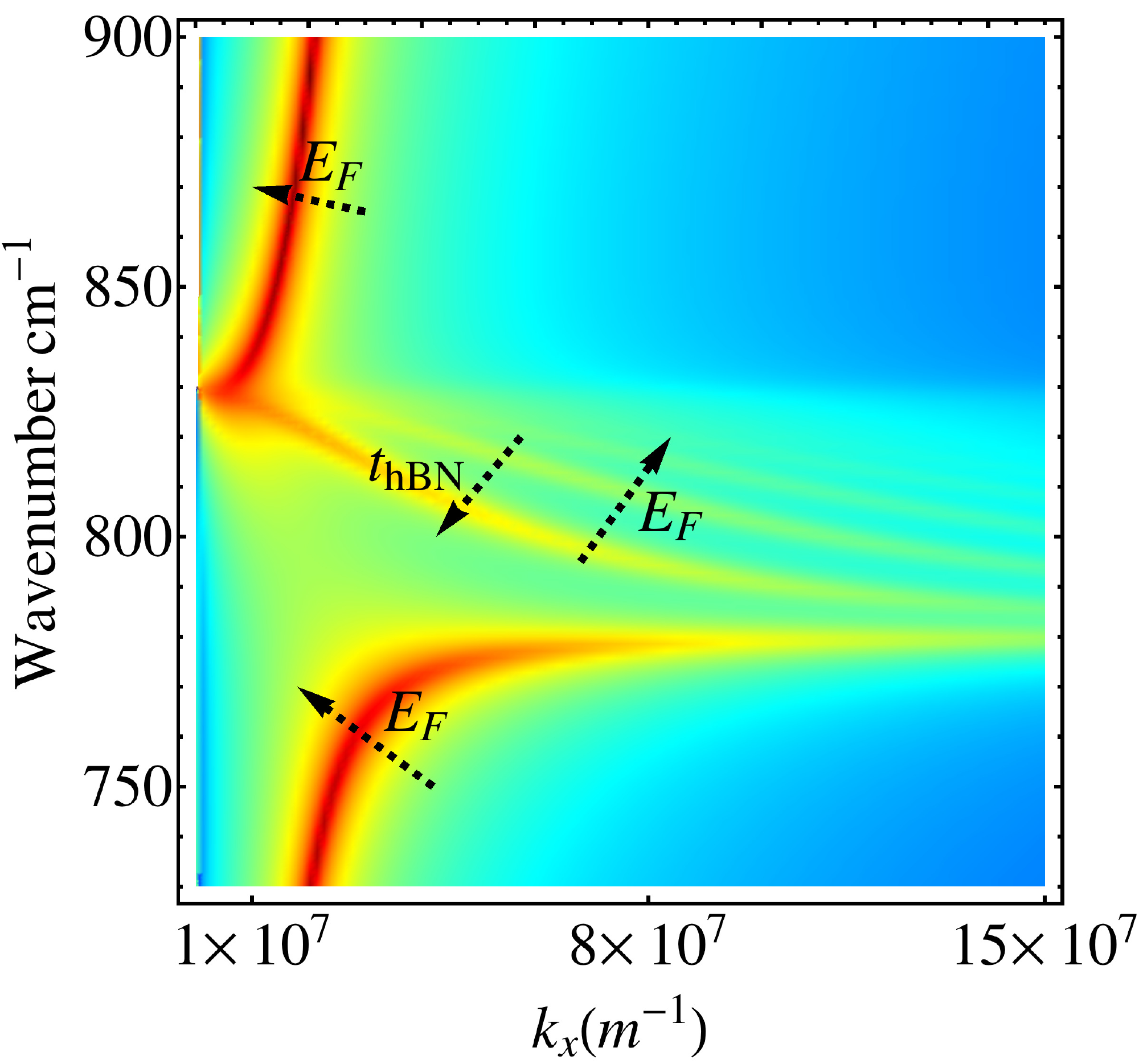}
\caption{Dispersion trends: out of plane}
\label{fig:TRENDS_OUT_OF_PLANE}
\end{subfigure}\\

\begin{subfigure}[b]{2.50in}
  \centering
  \includegraphics[width=2.50in]{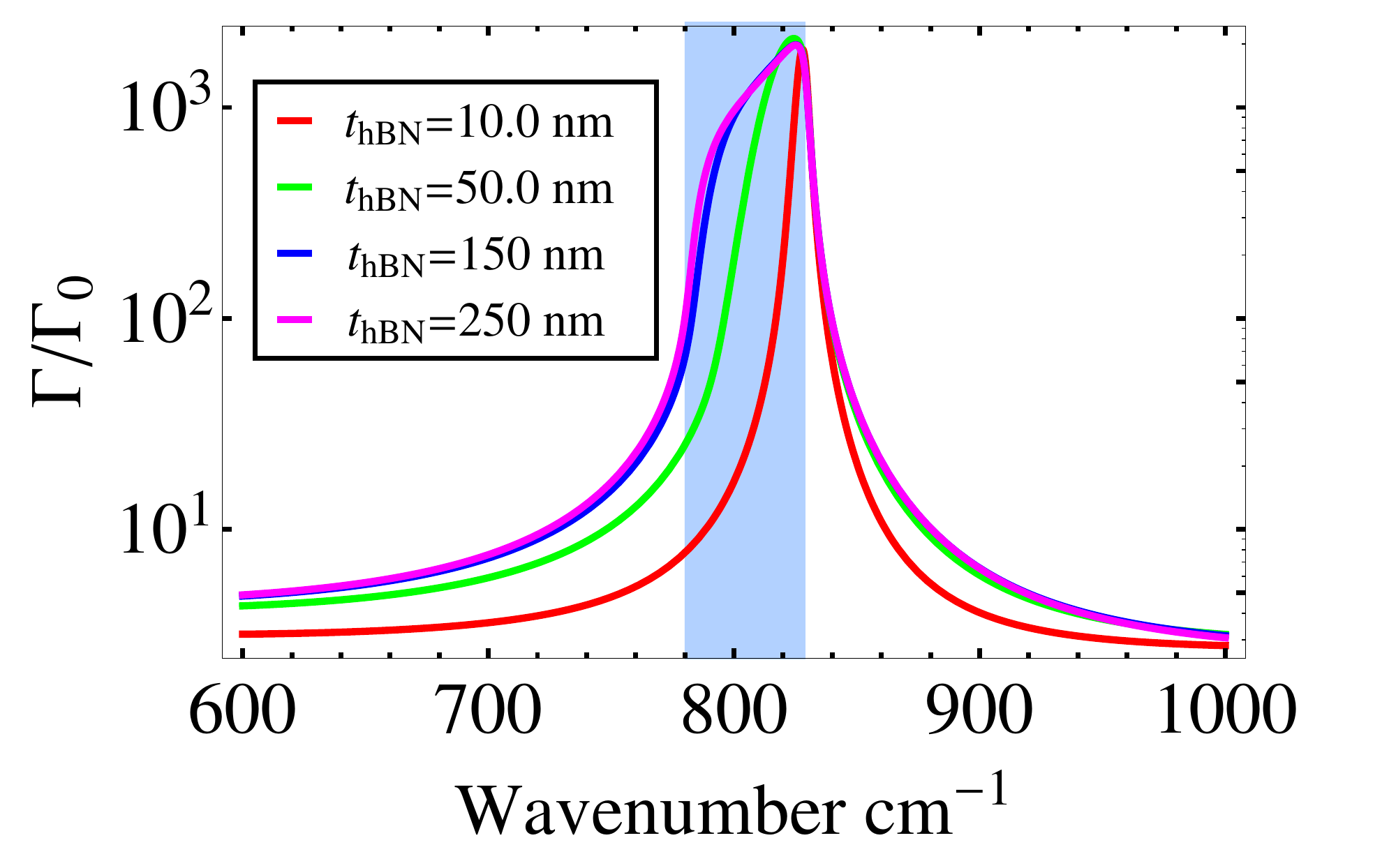}
  \caption{AHS}
  \label{fig:Purcell_thBN_dependence_out_of_plane_AHS}
\end{subfigure} \\

\begin{subfigure}[b]{2.25in}
  \centering
  \includegraphics[width=2.25in]{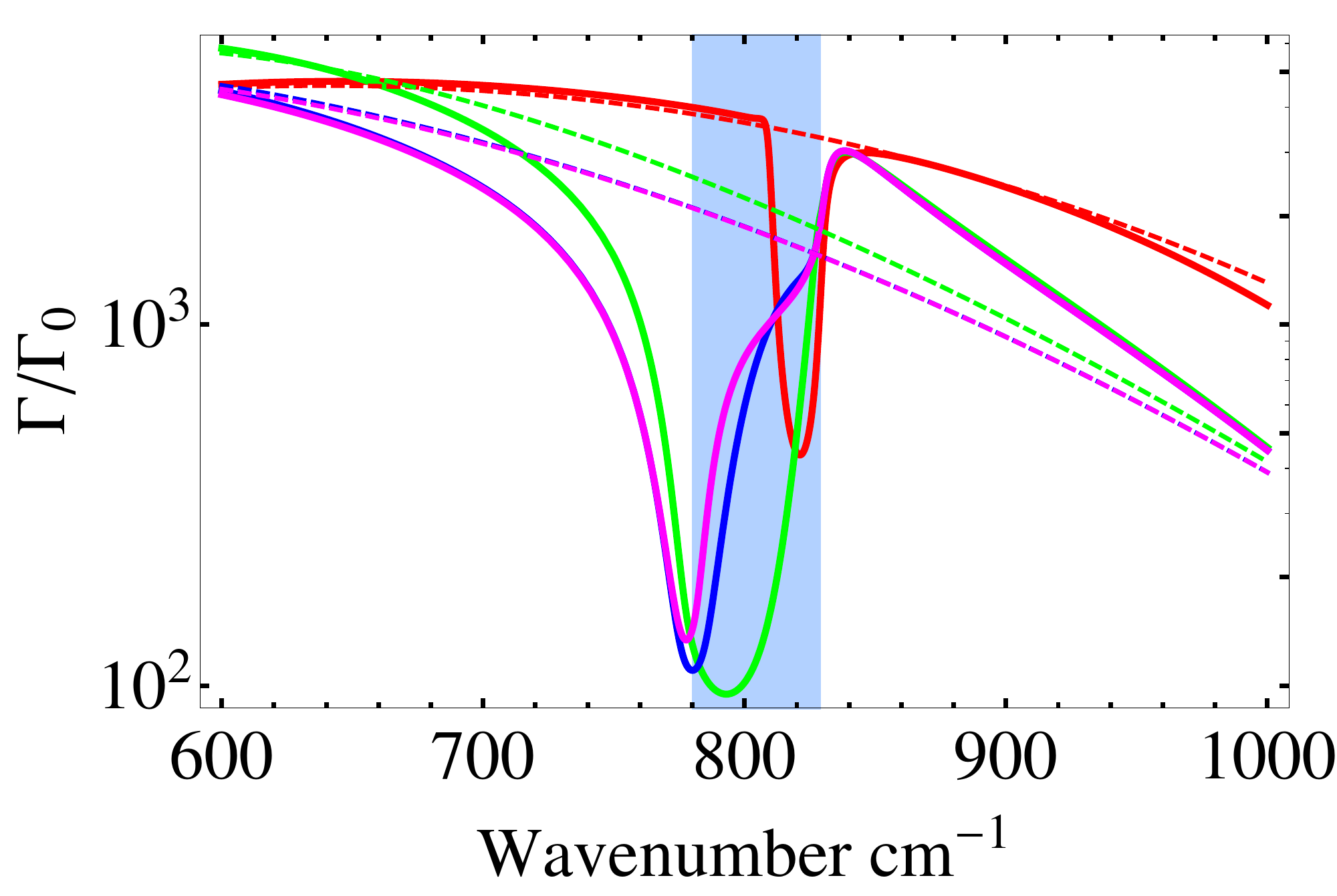}
  \caption{AGHS}
  \label{fig:Purcell_thBN_dependence_out_of_plane_AGHS}
\end{subfigure} 
\end{subfigure} 
\begin{subfigure}{3.2in} 
\begin{subfigure}[b]{2.2in}
\centering
\includegraphics[width=2.2in]{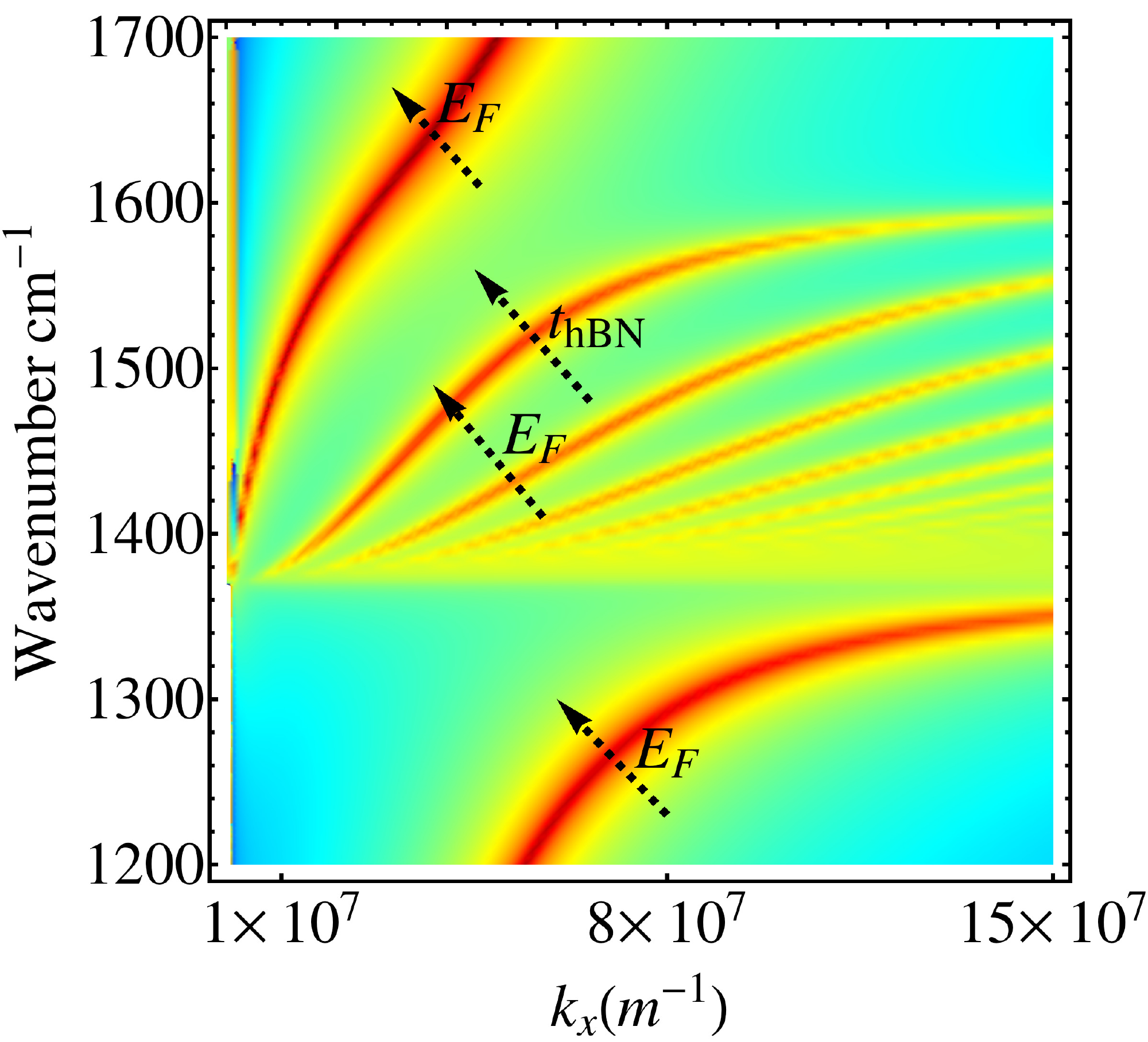}
\caption{Dispersion trends: in plane}
\label{fig:TRENDS_IN_PLANE}
\end{subfigure}\\

\begin{subfigure}[b]{2.25in}
  \centering
  \includegraphics[width=2.25in]{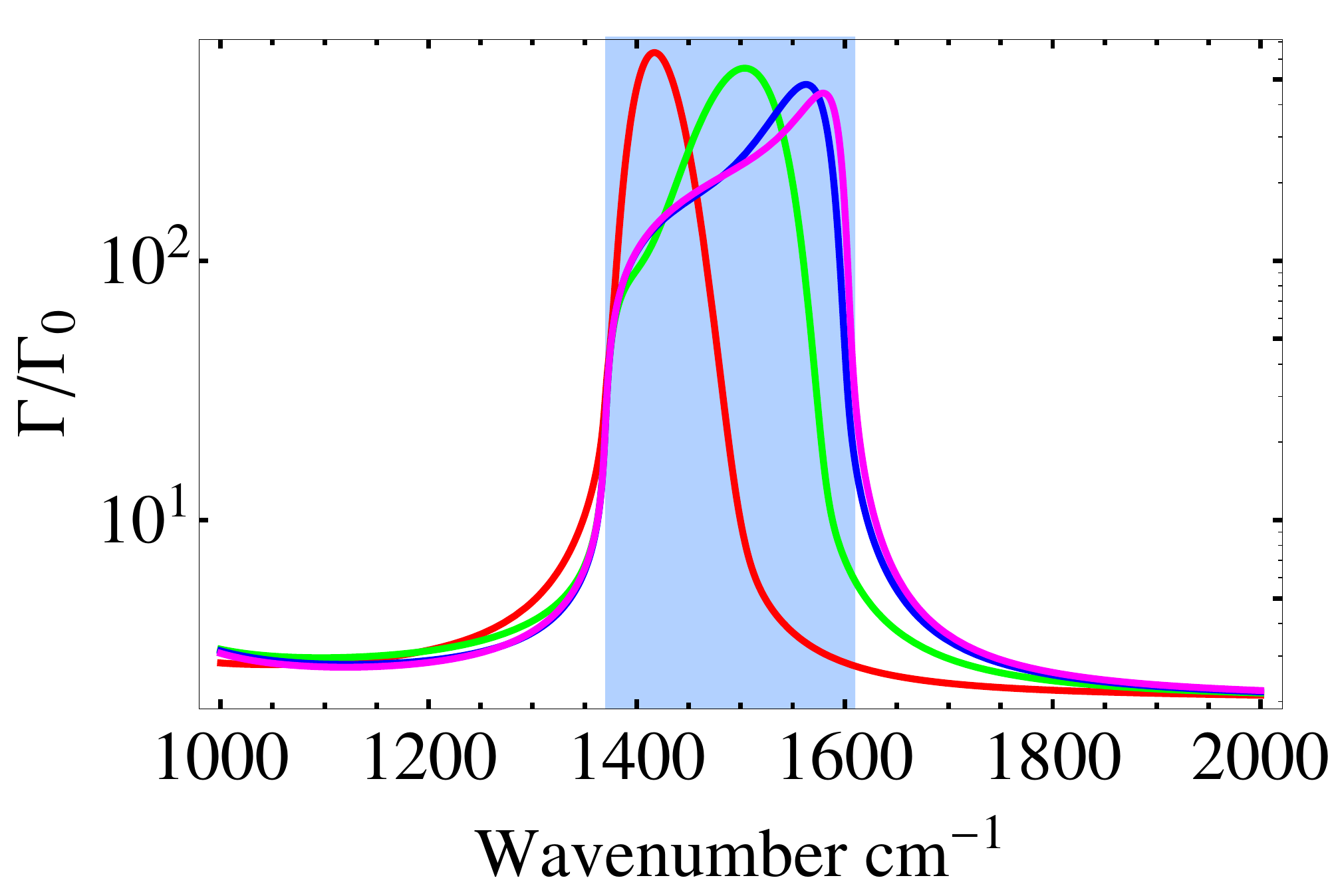}
  \caption{AHS}
  \label{fig:Purcell_thBN_dependence_in_plane_AHS}
\end{subfigure}\\

\begin{subfigure}[b]{2.25in}
  \centering
  \includegraphics[width=2.25in]{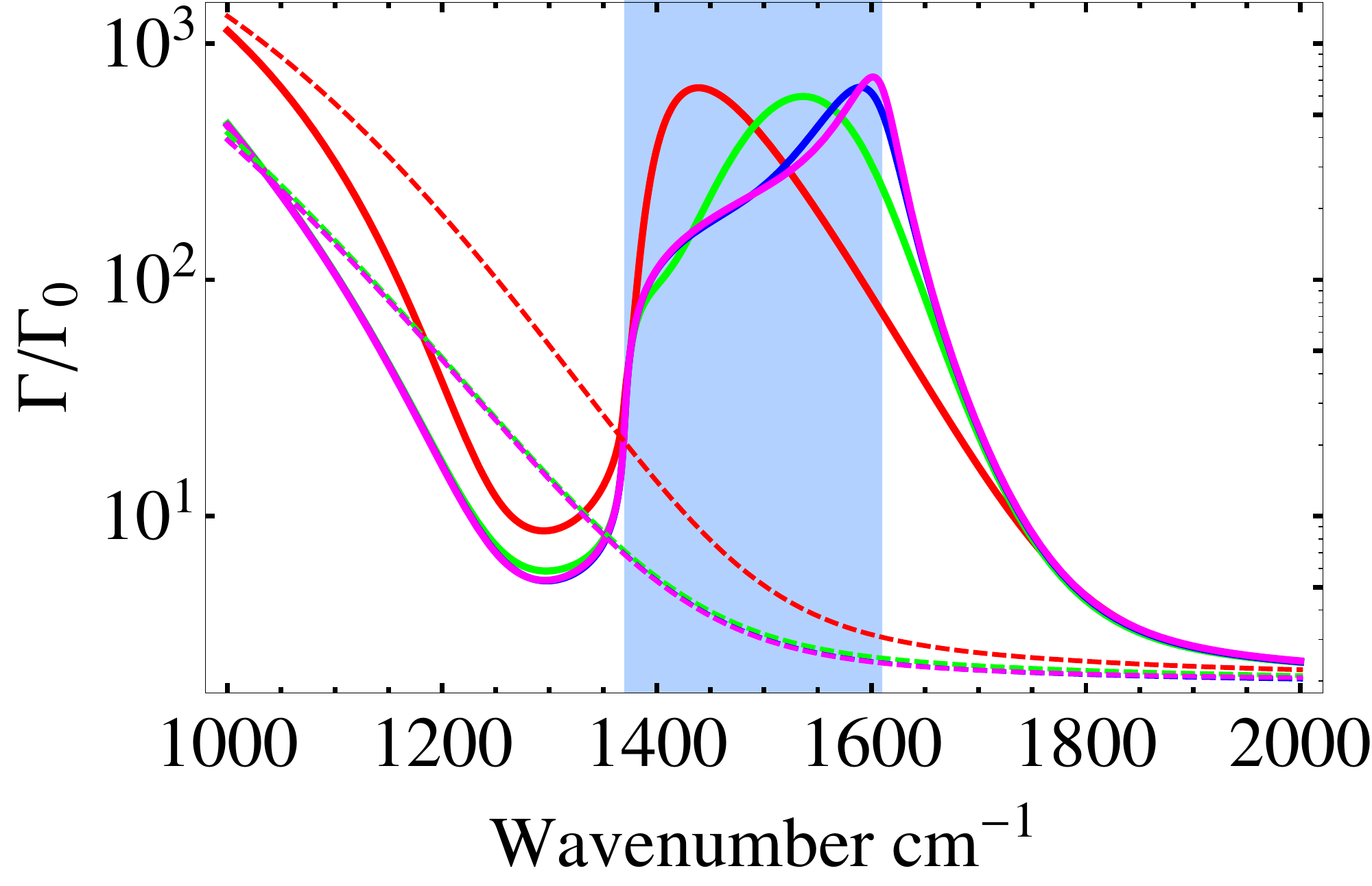}
  \caption{AGHS}
  \label{fig:Purcell_thBN_dependence_in_plane_AGHS}
\end{subfigure}
\end{subfigure}
\caption{ {\it hBN thickness dependent Purcell spectra in graphene-hBN system:}  {(a) and (d) represent trends for dispersion variation as a function of hBN thickness and graphene doping near the out of plane and in-plane phonon resonances.} We consider here the AGHS system. The arrows point in the direction of increasing $t_{hBN}$ or $E_F$. The trends shown are for independent variation of the two variables and not simultaneous. Unlike other dispersion plots in this paper, here the $x$-axis represents $k_x$ to make it easier for the reader to relate it to $1/d_s$. See text for details. { (b), (c), (e) and (f) represent hBN thickness dependent Purcell spectra in graphene-hBN system.} Distance $d_s$ of the quantum emitter from the hBN is fixed at $\SI{100}{nm}$. In the AGHS case, the emitter is on the graphene side. Dashed lines show the Purcell enhancement due to graphene plasmon without the phonon-polariton contribution from the hBN slab, that is, assuming $\epsilon_{hBN}(\omega\rightarrow 0)$.  Graphene Fermi level is assumed to be $\SI{0.5}{eV}$. }
\label{fig:Purcell_thBN_dependence}
\end{figure*}

The dispersion of guided modes in the hBN slab depends strongly on its thickness. This variation in the dispersion has a profound influence on the PLDOS spectra. Let's focus on the AHS out of plane phonon case first. In this case, as shown in Fig.~\ref{fig:TRENDS_OUT_OF_PLANE}, with increase in hBN thickness $t_{hBN}$, the phonon-polariton dispersion moves closer to $\omega_{TO}$ or the lower end of the RS band at small $k_x$. The implication of this movement is that the PLDOS peak starts moving from the upper end of the RS band (at small $t_{hBN}$), broadening toward the lower end (at large $t_{hBN}$). The broadening occurs again because as $t_{hBN}$ is increased, 1) more states are available throughout the RS band (at fixed $k_x$) and 2) the redshifted phonon polaritons, specially at low $k_x$, have greater curvature. Both of these expectations are verified in the calculations shown in Fig.~\ref{fig:Purcell_thBN_dependence_out_of_plane_AHS}.

With the graphene coating on the hBN slab, there is a spectral dip immediately below $\omega_{TO}$. As explained for the $d_s$ variation earlier, this is due to the broad peak arising from the curved nature of graphene dispersion at the chosen $d_s$, which is peaked much below $\omega_{TO}$. As $t_{hBN}$ increases, the width of the dip becomes narrow because it is ``pushed" from the right by the phonon-polariton states which are being red-shifted. This explains the movement of the spectral dip in Fig.~\ref{fig:Purcell_thBN_dependence_out_of_plane_AGHS}.

This behaviour persists for the in-plane modes as shown in Figs.~\ref{fig:Purcell_thBN_dependence_in_plane_AHS} and \ref{fig:Purcell_thBN_dependence_in_plane_AGHS}, with the difference being the opposite trend of peak movements, on account of differing types of hyperbolicity, as explained earlier.

\subsection{Active control of spectral dips}

\begin{figure*}
\centering
\begin{subfigure}{3.25in}
  \centering
  \includegraphics[width=3.25in]{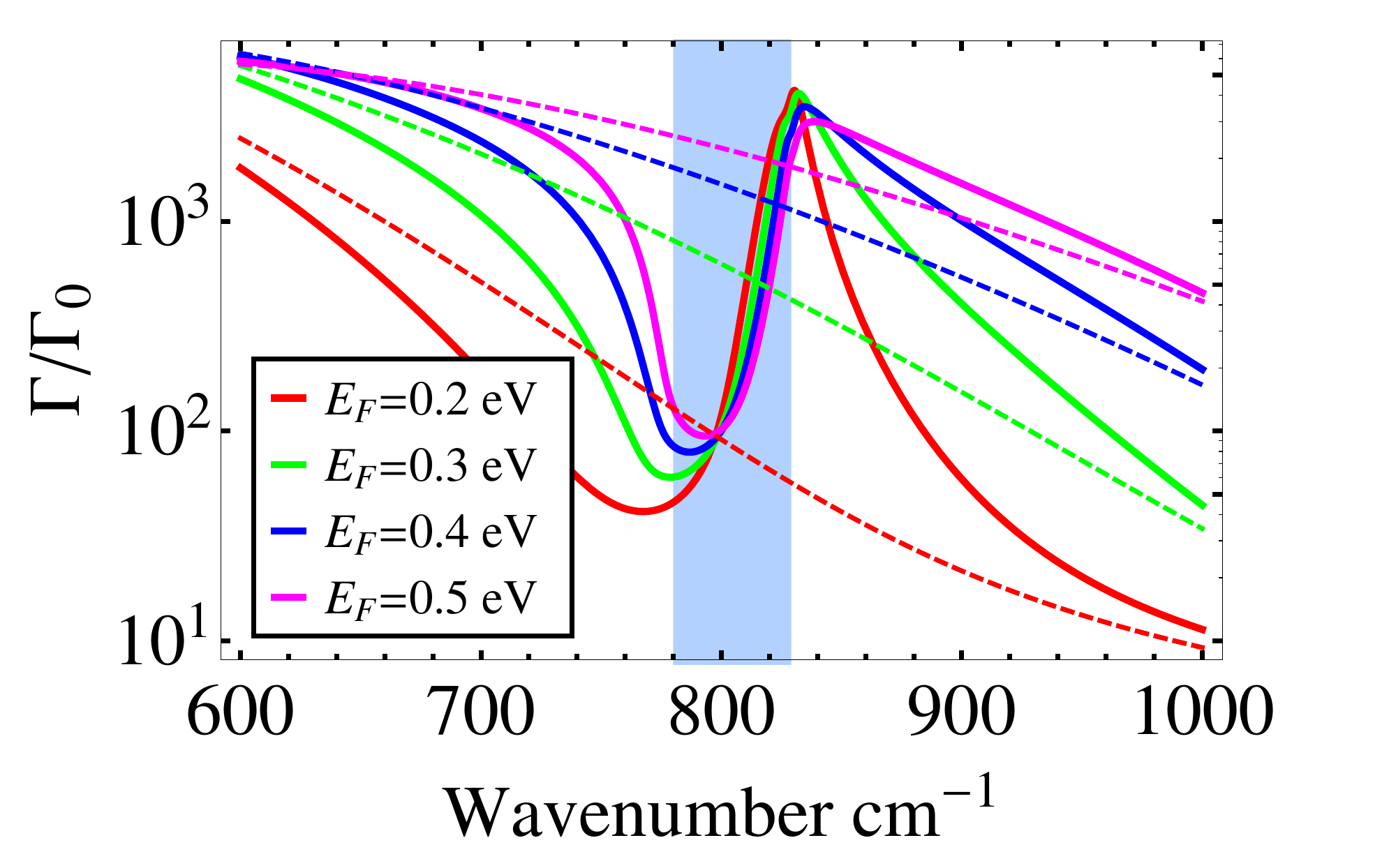}
  \caption{AGHS}
  \label{fig:Purcell_Ef_dependence_out_of_plane_AGHS_ds_100nm_thBN_50nm}
\end{subfigure} 
\begin{subfigure}{3in}
  \centering
\includegraphics[width=3in]{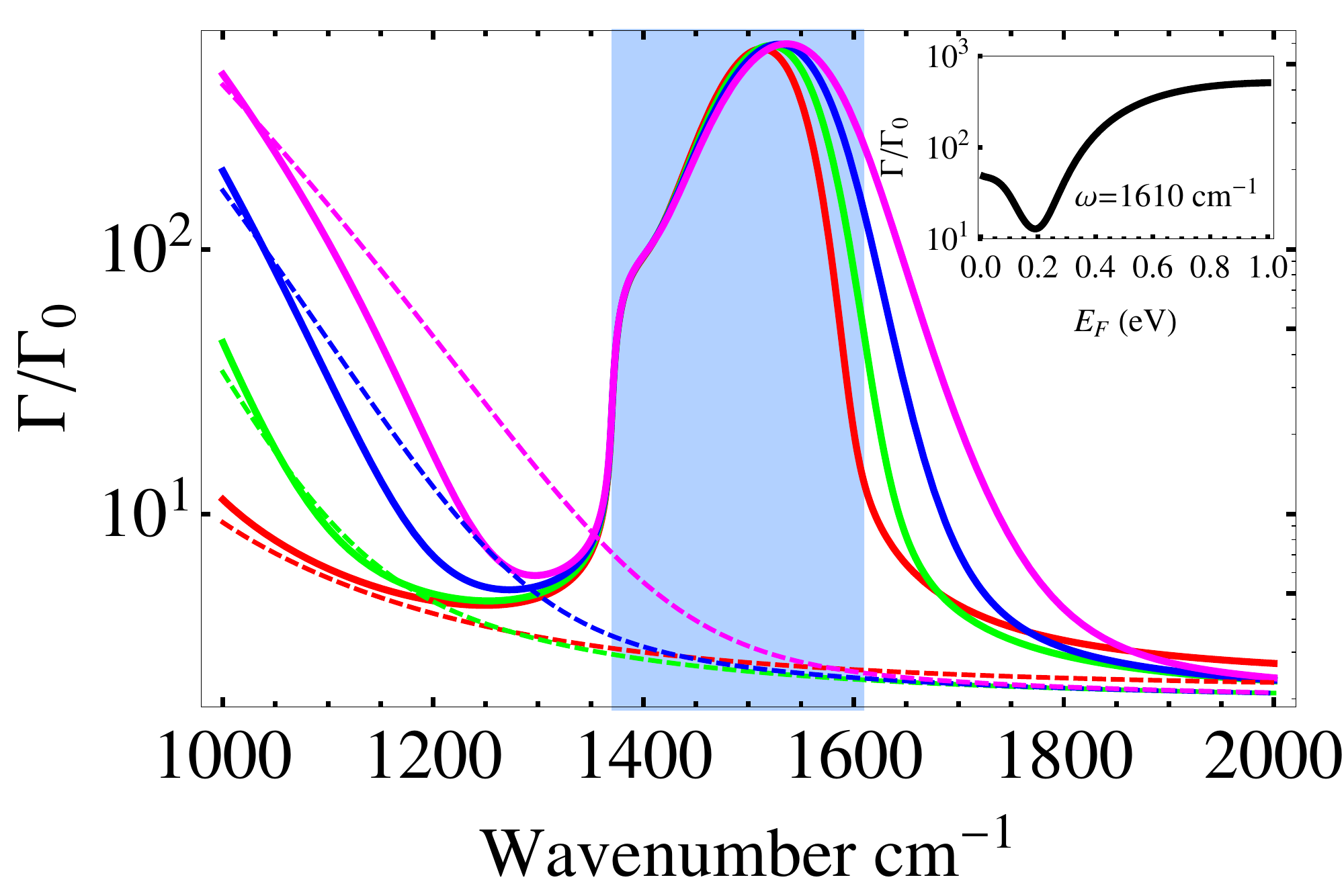}
  \caption{AGHS}
  \label{fig:Purcell_Ef_dependence_in_plane_AGHS_ds_100nm_thBN_50nm}
\end{subfigure}
\caption{{\it Fermi level dependent Purcell spectra in graphene-hBN system: } The doping of the graphene is varied. The thickness of the hBN slab is $\SI{50}{nm}$. Distance $d_s$ of the quantum emitter from the hBN is fixed at $\SI{100}{nm}$. The AGHS case has been considered with the emitter is on the air side. Subfigure (a) represents Purcell spectra around out-of-plane resonance and (b) represents that near in-plane resonance. Dashed lines show the Purcell enhancement due to graphene plasmon without the phonon-polariton contribution from the hBN slab, that is, assuming $\epsilon_{hBN}(\omega\rightarrow 0)$. {The inset in subfigure (b) presents the decay rate near $\omega=1610$cm$^{-1}$ as a function of graphene doping.}}
\label{fig:Purcell_Ef_dependence}
\end{figure*}

The Fermi level of graphene can be tuned via electrostatic doping\cite{10.1126/science.1102896}. This is one of the main motivations for using graphene for photonics applications\cite{10.1038/nnano.2014.215,10.1038/nnano.2011.146}. Importantly, graphene with higher doping has a plasmon dispersion which is closer to the light line\cite{doi:10.1021/nl201771h,  doi:10.1021/nn406627u, doi:10.1021/nl3047943}. Thus for a fixed dipole distance and hBN thickness, changing the Fermi level offers a route for active control of the spectral location and width of the spectral dips in PLDOS.

To demonstrate such a gate tunability of the PLDOS, we carried out calculations for the AGHS case as shown in Fig.~\ref{fig:Purcell_Ef_dependence}. For the out of plane case, we expect that at larger doping (or larger $E_F$), the graphene plasmon below $\omega_{TO}$ will flatten out at smaller $k_x$, as shown in Fig.~\ref{fig:TRENDS_OUT_OF_PLANE}. Conversely, at a fixed value of $d_s$, this graphene plasmon branch will have more curvature and will be farther below $\omega_{TO}$ for smaller doping. On the other hand, due to availability of states inside the RS band, the PLDOS is expected to rise there. This results in a dip occurring between the two peaks near $\omega_{TO}$.

Secondly, towards the upper end of this (out of plane) RS band, we observe an interesting trend. The half line width of the part of PLDOS peak which is inside the RS band decreases with increasing $E_F$. On the other hand, the line width of the PLDOS peak outside the RS band (at the higher frequency end), increases with increasing $E_F$. There are two effects in play here. Firstly, for smaller Fermi-level, since the graphene plasmon is very highly confined to the graphene itself, it does not interact very strongly with the phonon-polariton. As the doping is increased, the plasmon confinement decreases and as shown in Fig.~\ref{fig:TRENDS_OUT_OF_PLANE}, the phonon-polaritons inside the RS are repelled. At reasonably small $k_x$, not only does this repulsion cause a blue shift of the phonon-polaritons but also reduces the curvature, resulting in narrower linewidth inside the RS band at high graphene doping. For frequencies above $\omega_{LO}$, the graphene plasmon is flatter for smaller doping. This causes the linewidths of the PLDOS peaks to the right of $\omega_{LO}$ to broaden as doping is increased. A subtle point to note here is that the graphene damping is expected to decrease with increasing $E_F$ since we assumed the relaxation time $\tau$ to be proportional to $E_F$. But we have also checked that these trends persist for the case of constant $\tau$.
{
Similar but complementary trends are observed for the in-plane modes as shown in Fig.~\ref{fig:Purcell_Ef_dependence_in_plane_AGHS_ds_100nm_thBN_50nm}. Note that due to high frequency, the interband excitations begin to damp the graphene plasmon mode, which results in an overall decrease in the PLDOS of the uncoupled graphene itself. Interestingly, for the in-plane band, the spontaneous emission rate is most sensitive to graphene doping near $\omega_{LO}$ (1610cm$^{-1}$). On the other hand, further inside this RS band at lower frequencies, the decay rate is almost independent of the graphene Fermi level. Therefore, to study the emission near $\omega_{LO}$, in the inset of Fig.~\ref{fig:Purcell_Ef_dependence_in_plane_AGHS_ds_100nm_thBN_50nm}, we have plotted the decay rate at $\omega_{LO}$, as a function of the graphene Fermi level. First of all, we observe that as we go from higher to lower doping, the decay rate becomes smaller. This trend can be explained as before, using Fig.~\ref{fig:TRENDS_IN_PLANE}. In accordance with Fig.~\ref{fig:TRENDS_IN_PLANE}, with decreasing $E_F$, the phonon polariton plasmon modes redshift, resulting in lowering of the decay rate at this frequency (note that we have to look at constant $k_x=1/d_s$). Conversely, one can say that for this emitter location, higher gate voltage gives higher decay rate which arises due to a larger contribution from the phonon polariton modes.

 However, the above picture explains the trend only upto a certain doping. As shown in the inset of Fig.~\ref{fig:Purcell_Ef_dependence_in_plane_AGHS_ds_100nm_thBN_50nm}, below a certain Fermi level, the decay rate starts rising again and achieves a constant value corresponding to undoped graphene. This critical value occurs near $E_F\approx\omega$, which in this case happens to be around $E_F=\SI{0.2}{eV}$. To explain this, we note that as the frequency approaches the Fermi level, while the plasmon mode becomes weaker, the interband contribution becomes larger. We have checked that if we remove the interband damping term from the graphene conductivity, the decay rate continues to decrease as the Fermi level becomes smaller (see supplementary information). This suggests that the rise of the decay rate at small doping is explained by a decay route opening up through this lossy channel of interband transitions as we asymptotically approach the case of undoped graphene\cite{PhysRevB.84.165438,:/content/aip/journal/jcp/129/5/10.1063/1.2956498}. At zero temperature, this critical point should occur near $E_F\approx\omega/2$. However due to finite temperature smearing in the graphene conductivity, it is observed to occur near $E_F\approx\omega$. On decreasing the temperature we observed that the critical Fermi level indeed approaches $\omega/2$, as shown in the supplemental information.
 
Lastly, it should be noted that in this case, the shape of the decay rate versus $E_F$ (as shown in the inset of Fig.~\ref{fig:Purcell_Ef_dependence_in_plane_AGHS_ds_100nm_thBN_50nm}) is dependent on the emitter location ($d_s$). In general we expect there to be an optimum value of $E_F$ where the decay rate is maximized, corresponding to the plasmon-phonon dispersion coinciding at the given frequency $\omega$ and wavevector $k_x\approx1/d_s$. In the supplementary information, we further study the shape of this curve for different emitter locations.
}

\section{Conclusion and outlook}

In this work, we have shown that heterostructures comprised of graphene and hBN can provide a highly versatile tool for controlling light matter interaction at the nanoscale. The gate tunability of graphene plasmons can be combined with the high confinement property of hBN phonon polaritons, allowing for new metamaterials that marry the unique qualities of the two materials. Being a hyperbolic material, the hBN provides two completely different regimes of plasmon-phonon coupling in the presence of graphene. The spectral dips in spontaneous emission enhancement is an example of this coupling where the width of the dips are quite different between the two RS bands. Moreover, the contrast for this dip is more than an order of magnitude in both cases with the width reaching $20$ cm$^{-1}$ at a frequency of around $800$ cm$^{-1}$. The width of the spectral dip can be controlled by adjusting the Fermi level of graphene and the thickness of the hBN. This can provide a powerful route to turning the decay rate on and off via an external knob. In particular, this could find application in vibrational spectroscopy\cite{doi:10.1021/nl404824w} and stimulated Raman scattering, with the Purcell dip near $\omega_{TO}$ of either RS band effectively being used as a gate tunable notch filter. Our findings might find applications in photodetection\cite{10.1038/ncomms2951}, electrically tunable thermal management via broadband near field heat transfer\cite{doi:10.1021/ph5001633} and subwavelength imaging\cite{doi:10.1021/nn303845a, Taubner15092006}. The analysis and general principles presented in this paper apply equally well to other systems of hyperbolic materials, both natural\cite{10.1155/2012/267564} and artificial\cite{10.1155/2012/452502}. 

\begin{acknowledgement}
A.K. and N.X.F. acknowledge the financial support by the NSF (grant CMMI-1120724) and AFOSR MURI (Award No. FA9550-12-1-0488). T.L. acknowledges support from University of Minnesota. K.H.F. acknowledges financial support from Hong Kong RGC grant 509813. 
\end{acknowledgement}

\begin{suppinfo}
Detailed theoretical calculation method, dependence of Purcell spectra on various other combinations of graphene Fermi level, hBN thickness and distance of the emitter from the graphene.
\end{suppinfo}

\bibliography{hBN_G_final}

\end{document}